\documentclass[manuscript, screen]{acmart}

\usepackage{tikz}
\usepackage{amsmath}

\usepackage{filecontents}

\usepackage{enumitem}
\usepackage{xstring}
\newcommand{\PP}[1]{\noindent\textbf{#1}}

\newcommand{\sysname}{\textsc{DeepAVR}}

\PassOptionsToPackage{dvipsnames}{xcolor}
\usepackage{pifont} %
\usepackage{booktabs}
\usepackage{multirow}
\usepackage{listings}
\usepackage{adjustbox}
\usepackage{array}
\usepackage{makecell}
\usepackage{longtable}
\usepackage{xspace}
\usepackage{colortbl}
\usepackage{algorithm}
\usepackage{algpseudocode}

\usepackage{tcolorbox}

\PassOptionsToPackage{hyphens}{url} %
\usepackage{hyperref}
\let\oldding\ding %

\renewcommand{\ding}[1]{%
  \ifnum#1=55
    {\textcolor{red}{\oldding{#1}}}%
  \else
    \oldding{#1}%
  \fi
}

\definecolor{ForestGreen}{rgb}{0.13, 0.55, 0.13}
\renewcommand{\checkmark}{\textcolor{ForestGreen}{\ding{51}}}
\newcommand{\orangetriangle}{{\color{orange}\blacktriangle}}

\usepackage{graphicx}
\usepackage{subcaption}

\usepackage{listings}
\usepackage{xcolor}

\definecolor{dvAdd}{RGB}{0,110,40}      %
\definecolor{dvRem}{RGB}{185,25,25}     %
\definecolor{dvCmt}{RGB}{110,110,110}
\definecolor{dvFrame}{RGB}{200,200,200}
\definecolor{dvHdr}{RGB}{232,232,232}
\definecolor{dvBg}{RGB}{250,250,250}

\newcommand{\codehdr}[1]{%
  \par\noindent
  \colorbox{dvHdr}{%
    \makebox[\dimexpr\linewidth-2\fboxsep\relax][l]{%
      \rule[-0.55ex]{0pt}{2.05ex}\scriptsize\ttfamily #1}}%
  \par\nointerlineskip}

\newcommand{\textpanel}[1]{%
  \par\noindent
  \fcolorbox{dvFrame}{dvBg}{%
    \parbox{\dimexpr\linewidth-2\fboxsep-2\fboxrule\relax}{%
      \scriptsize\raggedright #1}}%
  \par}
\newcommand{\dvitem}[2]{%
  \par\hangindent=1.1em\hangafter=1\noindent
  \hspace*{0.3em}\makebox[0.8em][l]{\textbullet}\textbf{#1:} #2}

\lstdefinestyle{dvBase}{
  basicstyle=\scriptsize\ttfamily,
  numbers=none, frame=single, rulecolor=\color{dvFrame},
  backgroundcolor=\color{dvBg},
  columns=fullflexible, keepspaces=true, breaklines=false,
  xleftmargin=1pt, xrightmargin=1pt, framexleftmargin=1pt,
  aboveskip=1.9pt, belowskip=0pt, framesep=1.5pt, %
  moredelim=**[is][\color{dvRem}]{~R}{R~},
  moredelim=**[is][\color{dvAdd}]{~G}{G~},
}

\lstdefinestyle{dvReport}{style=dvBase}

\lstdefinestyle{dvCode}{language=C, style=dvBase,
  commentstyle=\color{dvCmt}\itshape, keywordstyle=\bfseries}

\lstdefinestyle{dvDiff}{language=C, style=dvBase,
  commentstyle=\color{dvCmt}\itshape, keywordstyle=\bfseries,
  morecomment=[f][\color{dvAdd}]{+},
  morecomment=[f][\color{dvRem}]{-}}

\AtBeginDocument{%
  }

\setcopyright{rightsretained}
\copyrightyear{2026}
\acmYear{2026}

\begin{document}

\title{DeepAVR: Deep Analysis via Multi-Agent Collaboration for Automated Vulnerability Repair}

\author{Woorim Han}
\authornote{These authors contributed equally to this work.}
\email{rimwoo98@snu.ac.kr}
\affiliation{%
  \institution{Seoul National University}
  \city{Seoul}
  \country{Republic of Korea}
}

\author{Hyunjun Kim}
\authornotemark[1]
\email{hjkim1@mju.ac.kr}
\affiliation{%
  \institution{Myongji University}
  \city{Yongin}
  \country{Republic of Korea}
}

\author{Brian Junekyu Lee}
\email{blee676@gatech.edu}
\affiliation{%
  \institution{Georgia Institute of Technology}
  \city{Atlanta}
  \country{United States}
}

\author{Kyeongmin Kim}
\email{kim1254@unist.ac.kr}
\affiliation{%
  \institution{Ulsan National Institute of Science and Technology (UNIST)}
  \city{Ulsan}
  \country{Republic of Korea}
}

\author{Younghan Lee}
\authornote{Corresponding authors.}
\email{yhlee@sungshin.ac.kr}
\affiliation{%
  \institution{Sungshin Women's University}
  \city{Seoul}
  \country{Republic of Korea}
}

\author{Hyungon Moon}
\email{hyungon@unist.ac.kr}
\affiliation{%
  \institution{Ulsan National Institute of Science and Technology (UNIST)}
  \city{Ulsan}
  \country{Republic of Korea}
}

\author{Yunheung Paek}
\authornotemark[2]
\email{ypaek@snu.ac.kr}
\affiliation{%
  \institution{Seoul National University}
  \city{Seoul}
  \country{Republic of Korea}
}

\renewcommand{\shortauthors}{Han et al.}

\begin{abstract} 

Automated Vulnerability Repair (AVR) has emerged as a promising approach to identifying and fixing software vulnerabilities autonomously. Early LLM-based approaches focused on patch generation under the assumption of perfect fault localization, while recent end-to-end methods attempt to repair vulnerabilities directly from sanitizer logs. However, these methods often fail when the root cause is deeply hidden—either \textit{physically} distant from the crash site or \textit{logically} masked by interdependent errors—making the initial report insufficient.
We propose \sysname{}, a multi-agent framework that transforms intermediate repair outcomes into structured diagnostic feedback. It coordinates specialized agents for fault localization, patch generation, compilation, security and functionality validation, and patch review to iteratively refine localization and patch hypotheses. Specifically, \sysname{} introduces patch-conditioned re-localization, using execution results from intermediate patches to uncover hidden root causes, and root-cause-oriented patch review to identify symptom-level fixes and incomplete repairs that escape execution-based validation.
Evaluation on VulnLoc, a real-world C/C++ vulnerability repair dataset, shows that \sysname{} outperforms prior methods, generating correct patches for 92.3\% of vulnerabilities. Evaluation on SAN2VULN demonstrates its ability to repair vulnerabilities unseen during training. On SEC-bench, \sysname{} outperforms the evaluated baselines on cases whose fixes lie outside the crash-site function.

\end{abstract}



\maketitle

\section{Introduction}
\label{s:Intro}

The inherent complexity of modern software has contributed to the continued prevalence of security flaws, commonly referred to as vulnerabilities.
This prevalence underscores the urgency of timely remediation of vulnerabilities.
However, manual repair by experts is time-consuming and labor-intensive, making it difficult to keep up with the rapid emergence of new vulnerabilities, which motivates the need for automated approaches.

\textit{Automated program repair} (APR) has received substantial attention as a promising direction to automatically address software defects~\cite{zhang2023gamma,ye2022selfapr,zhao2024repair,zhang2024autocoderover,meng2023template}. While early APR research largely focused on general bug fixing to restore functional correctness, subsequent work has recognized that repairing security vulnerabilities—often termed \textit{automated vulnerability repair} (AVR)~\cite{zhang2022program,extractfix,hong2020saver,crashrepair,gao2019crash}—demands more than restoring functional correctness; it must additionally eliminate exploitability. Whereas test-suite-based APR uses failing and passing tests to guide repair, end-to-end AVR often starts from a single proof-of-concept input and its sanitizer report, which provide only limited evidence of the underlying vulnerability.

More recently, large language models (LLMs) have emerged as a promising paradigm for AVR~\cite{icse24,fse22,shao2026fix,kim2025logs,yu2025patchagent,looprepair,AVR_ASE25_MUSSEL,AVR_ASE25_Vul-R2}. Pre-trained on vast corpora of code, these models possess the capacity to comprehend program semantics and synthesize code, facilitating the automated generation of patches for a wide spectrum of vulnerability types. When coupled with repair-oriented prompting and program context, LLMs can produce candidate fixes without relying on manually crafted rules, and their generalization capacity makes them attractive for repairing previously unseen patterns~\cite{sp23,sec24}.

Although early studies assumed perfect \textit{fault localization} (FL)~\cite{fse22,icse24,sp23,sec24, AVR_ASE25_MUSSEL, AVR_ASE25_Vul-R2}, 
recent end-to-end solutions generate patches directly from a single Proof-of-Concept (PoC)~\cite{looprepair} or the resulting sanitizer report~\cite{yu2025patchagent,kim2025logs}, a common form of vulnerability reporting.
These methods move AVR closer to practical settings by integrating localization, patch generation, and validation into a single pipeline. For example, LoopRepair~\cite{looprepair} combines CrashAnalysis~\cite{crashrepair}-based localization with iterative refinement, while SAN2PATCH~\cite{kim2025logs} and PatchAgent~\cite{yu2025patchagent} use LLM-based reasoning and tool-augmented interaction to repair vulnerabilities directly from sanitizer logs and retrieved code context. 
However, despite these advances, existing methods still struggle when the initial report provides only weak clues about the true root cause.
Such weak signals commonly arise in two forms: \textit{physical distance}, where the crash occurs far downstream from the silent root cause, and \textit{logical distance}, where the reported error message captures only a symptom rather than the initiating defect. In these cases, patch generation can be misled by the immediate crash context, causing existing methods either to miss the underlying fault condition or to produce superficial fixes that silence the crash without repairing the underlying vulnerability.

We observe that this limitation largely stems from the fact that existing methods fall short of fully diagnostic iterative repair. In PatchAgent~\cite{yu2025patchagent} and SAN2PATCH~\cite{kim2025logs}, validation is used primarily as a gate: it checks whether PoC execution and regression tests pass, but rarely exploits the specific nature of failures to revise the underlying root-cause hypothesis. LoopRepair~\cite{looprepair} incorporates richer feedback through taint-trace-guided selection, yet it still refines patches largely within the repair locations and candidate space defined by its initial pipeline. As a result, when a candidate patch suppresses a crash through a symptom-level fix (e.g., local guards or early exits), these systems often lack a mechanism to recognize the deeper logical deficiency and steer repair toward a developer-level fix.

In this paper, we present \sysname{}, a multi-agent framework for end-to-end AVR. Unlike prior iterative solutions, \sysname{} handles weak signals through \textit{deep analysis}, transforming intermediate repair outcomes into structured diagnostic feedback through three key design mechanisms.
First, \sysname{} adopts \textit{failure-aware role decomposition and structured feedback routing}. It organizes repair into specialized agents for fault localization, patch generation, compilation, security, functionality validation, and patch review. Rather than repeatedly generating patches without diagnosing failures, each iteration produces structured diagnostic feedback from validation agents, which is routed to refine either localization or patch generation. This structured collaboration allows the system to progressively shift from crash symptoms toward root-cause fixes, even when initial signals are weak due to physical or logical distance from the defect.

Second, \sysname{} introduces \textit{patch-conditioned re-localization} to convert intermediate failures into stronger diagnostic signals. Applying a candidate patch can expose deeper defects by altering the type or location of the observed failure. \sysname{} re-invokes the Fault Localization Agent on these updated reports, treating them as informative probes rather than noise. This enables the system to uncover root-cause candidates that are initially obscured, in contrast to prior approaches that either discard patches exhibiting new failure signals~\cite{looprepair} or fail to systematically exploit execution feedback obtained after applying intermediate patches~\cite{yu2025patchagent}.

Third, \sysname{} incorporates a \textit{root-cause-oriented patch review} stage that complements execution-based validation. The Patch Review Agent identifies symptom-level fixes (e.g., guard-only edits), detects misplaced validation logic (e.g., applying checks in the callee instead of enforcing preconditions at the caller), and ensures patch completeness across logically equivalent sites. By providing targeted feedback on repair quality, this stage pushes patches toward root-cause-correct solutions beyond what PoC and regression testing alone can guarantee.

We evaluate \sysname{} on three real-world benchmarks under standard end-to-end AVR settings~\cite{looprepair,kim2025logs}. First, we evaluate \sysname{} on VulnLoc~\cite{shen2021localizing}, the standard benchmark used in prior end-to-end AVR work~\cite{kim2025logs,crashrepair,looprepair}, enabling direct comparison. Out of 39 real-world vulnerabilities, \sysname{} generates plausible patches for all 39 cases and correct patches for 36, substantially outperforming prior baselines~\cite{kim2025logs,looprepair,crashrepair,specrover,yu2025patchagent}.
Next, we evaluate \sysname{} on SAN2VULN~\cite{kim2025logs}, specifically targeting vulnerabilities absent from the base model’s training set to verify its generalization to unseen cases. \sysname{} successfully generates correct patches for 22 out of 25 vulnerabilities, achieving an 88\% success rate. We further evaluate \sysname{} on the 200-CVE evaluation split of SEC-bench~\cite{secbench} to assess its ability to repair vulnerabilities whose fixes lie away from the crash site.
With DeepSeek-V4.1-Flash, \sysname{} resolves 111 (99.1\%) of the 112 instances whose developer fixes do not modify the crash-site function and achieves the highest function-level localization coverage among the evaluated methods. Finally, we perform ablation studies to measure the contribution of each specialized agent, and analyze the token costs and total runtime of \sysname{}. In summary, our key contributions are as follows:
\begin{itemize}[leftmargin=*]
\item \textbf{Failure-aware role decomposition and structured feedback routing.}
We present \sysname{}, which performs deep analysis by transforming intermediate repair outcomes into structured diagnostic feedback and routing them to the most appropriate repair stage, enabling effective repair even with weak sanitizer signals.

\item \textbf{Patch-conditioned re-localization.}
We introduce an iterative repair mechanism that leverages execution feedback from intermediate patches to unmask deeply hidden diagnostic signals and refine root-cause hypotheses.

\item \textbf{Root-cause-oriented patch review.}
We design a Patch Review Agent that identifies symptom-level fixes and incomplete repairs, enforcing root-cause-oriented patch correctness beyond execution-based validation.

\end{itemize}

The source code of DeepAVR is publicly available at
\url{https://github.com/rimwoohan/DeepAVR}.

\section{Motivating Example}
\label{s:ME}

\begin{figure}[ht!]
    \centering
    \begin{subfigure}[t]{0.49\linewidth}
        \codehdr{AddressSanitizer report}
\begin{lstlisting}[style=dvReport]
~R==ERROR: AddressSanitizer: heap-use-after-freeR~
READ of size 8 at 0x6020000000d0 thread T0
  #0 getName                decompile.c:398
  #1 decompileDELETE        decompile.c:3039
  ...
freed by thread T0 here:
  #0 realloc                (libasan.so.4)
  #1 ~RparseSWF_ACTIONRECORD  parser.c:1008R~
  ...
previously allocated by thread T0 here:
  #0 __interceptor_calloc   (libasan.so.4)
  #1 ~RparseSWF_ACTIONRECORD  parser.c:971R~
~RSUMMARY: heap-use-after-freeR~
~R         decompile.c:398 in getNameR~
\end{lstlisting}
        \captionsetup{margin=0.3em}
        \caption{AddressSanitizer report for CVE-2018-8964 (heap-use-after-free).}
        \label{subfig:san_8964}
    \end{subfigure}
    \hfill
    \begin{subfigure}[t]{0.49\linewidth}
        \codehdr{/src/util/parser.c}
\begin{lstlisting}[style=dvCode]
 966 case SWFACTION_PUSH: {
 967   int end;
 968   struct SWF_ACTIONPUSHPARAM *param;
 970   end = fileOffset + act->Length;
 971   ~Ract->Params = calloc(1, sizeof(*param));R~
 972   act->NumParam = 0;
 974   while (fileOffset < end) {
 975     param = &(act->Params[act->NumParam++]);
 ...      /* parse one param into *param */
1008     ~Ract->Params = realloc(act->Params,R~
           ~R(act->NumParam + 1) * sizeof(...));R~
1009   }
1010   break;
1011 }
\end{lstlisting}
        \captionsetup{margin=0.3em}
        \caption{Agent’s initial patch location derived from the crash report.}
        \label{subfig:parser_8964}
    \end{subfigure}

    \vspace{0.6em}

    \begin{subfigure}[t]{0.49\linewidth}
        \codehdr{/src/util/decompile.c}
\begin{lstlisting}[style=dvDiff]
 case PUSH_CONSTANT:   /* CONSTANT8 */
+ if (act->p.Constant8 > poolcounter) {
+   SWF_warn("WARNING: retrieving constants "
+            "not present in the pool.\n");
+   break;
+ }
 t = malloc(strlenext(pool[act->p.Constant8])+1);
 strcpyext(t, pool[act->p.Constant8]);

 case PUSH_CONSTANT16: /* CONSTANT16 */
+ if (act->p.Constant16 > poolcounter) {
+   /* ... the same guard ... */
+ }
 t = malloc(strlenext(pool[act->p.Constant16])+1);
\end{lstlisting}
        \captionsetup{margin=0.3em}
        \caption{Developer fix: bounds-check constant-pool access in decompile.c.}
        \label{subfig:dev_8964}
    \end{subfigure}
    \hfill
    \begin{subfigure}[t]{0.49\linewidth}
        \codehdr{/src/util/decompile.c}
\begin{lstlisting}[style=dvDiff]
 case PUSH_CONSTANT:   /* CONSTANT8 */
+ if (!pool || act->p.Constant8 < 0 ||
+     act->p.Constant8 >= pool_size ||
+     !pool[act->p.Constant8]) return "this";
 t = malloc(strlenext(pool[act->p.Constant8])+1);
 strcpyext(t, pool[act->p.Constant8]);
 if (strlen(t))  /* not a zero length string */

 case PUSH_CONSTANT16: /* CONSTANT16 */
+ if (!pool || act->p.Constant16 < 0 ||
+     act->p.Constant16 >= pool_size ||
+     !pool[act->p.Constant16]) return "this";
 t = malloc(strlenext(pool[act->p.Constant16])+1);
 strcpyext(t, pool[act->p.Constant16]);
\end{lstlisting}
        \captionsetup{margin=0.3em}
        \caption{\sysname's patch after patch-conditioned re-localization.}
        \label{subfig:our_8964}
    \end{subfigure}

    \caption{Logically distant vulnerability example (CVE-2018-8964 in libming, heap use-after-free).}
    \label{Fig:motivation}
\end{figure}

A sanitizer report is the primary, and often the only, evidence available for repair, yet it records where a program failed rather than why. As introduced in \autoref{s:Intro}, this gap takes two forms. Under \textit{logical distance}, the reported error message captures only a symptom rather than the initiating defect. Under \textit{physical distance}, the crash occurs far downstream from a silent root cause. 
We use one example of each: in the logically distant case, an intermediate patch changes the observed failure, providing new evidence for localization; in the physically distant case, a symptom-level patch passes the available execution-based checks but leaves the underlying defect unresolved.

\subsection{Logically Distant Vulnerability (CVE-2018-8964)}
\label{ss:logical}
The AddressSanitizer (ASan) report (\autoref{subfig:san_8964}) labels the failure as a heap use-after-free (UAF) and points to a read in \texttt{decompile.c}. At the same time, it attributes the heap object’s allocation and resizing to \texttt{parseSWF\_ACTIONRECORD} in \texttt{parser.c}. 

\PP{Parser-side allocation and growth.}
Since the sanitizer reports the failure as a UAF and highlights the allocation/free lifecycle, a repair agent is naturally drawn to the parser-side memory management code.
In \texttt{parser.c} (\autoref{subfig:parser_8964}), the \texttt{SWFACTION\_PUSH} handler allocates \texttt{act->Params} via \texttt{calloc} (line~971) and grows it via \texttt{realloc} (line~1008).
Each iteration takes an element address, \texttt{param = \&(act->Params[act->NumParam++])} (line~975), writes through \texttt{*param}, and may then resize the underlying buffer.
Viewed in isolation, this pattern suggests a plausible UAF hypothesis: if \texttt{realloc} moves the array, a previously taken element pointer could become dangling.

\PP{Decompiler-side dereference and developer fix.}
However, the true repair target lies in \texttt{decompile.c}, not in the parser-side allocation logic.
As shown in \autoref{subfig:dev_8964}, the decompiler performs constant-pool accesses such as \texttt{pool[act->p.Constant8]} and \texttt{pool[act->p.Constant16]}.
If these indices are out of range, the lookup can return an invalid pointer that is later dereferenced, eventually surfacing as the UAF reported by ASan.
The developer patch, therefore, adds bounds checks before the pool accesses, directly addressing the root cause.

\PP{Developer-aligned repair via patch-conditioned re-localization.}
Here, the reported UAF is a downstream symptom, not the true repair target. 
The initial parser-side patch replaces repeated reallocations with a single allocation after parameter parsing. 
Re-executing the same PoC on the patched program changes the observed failure from a heap use-after-free to a NULL-pointer dereference in \texttt{strlenext}, called by \texttt{getName} in \texttt{decompile.c}. This updated report provides a concrete diagnostic cue to revisit the constant-pool access at the call site.
Prior systems handle such pivots weakly: SAN2PATCH~\cite{kim2025logs} simply retries, PatchAgent~\cite{yu2025patchagent} reasons over the new report without applying the triggering patch, and LoopRepair~\cite{looprepair} discards type-shifting candidates. In contrast, \sysname{} applies the triggering patch and re-localizes on the resulting sanitizer report, treating the type shift as diagnostic evidence to move repair from the symptom to the underlying defect (\autoref{subfig:our_8964}).

\newsavebox{\physReportPanel}
\newsavebox{\physGuardPanel}
\newsavebox{\physReviewPanel}
\newsavebox{\physFixPanel}
\newlength{\physTopRowHeight}
\newlength{\physBottomRowHeight}
\begin{figure}[ht!]
    \centering
    \begin{lrbox}{\physReportPanel}
        \begin{minipage}[t]{0.48\linewidth}
        \vspace{0pt}
        \codehdr{AddressSanitizer report}
\begin{lstlisting}[style=dvReport]
~R==ERROR: AddressSanitizer: SEGV on unknownR~
~R  address 0x000000000000R~ (...)
The signal is caused by a READ memory access.
Hint: address points to the zero page.
  #0 ~RAcseConnection_parseMessageR~       (...)
  #1 IsoConnection_handleTcpConnection (...)
  #2 handleClientConnections     iso_server.c
  #3 handleIsoConnections        iso_server.c
  #4 IsoServer_processIncomingMessages (...)
  ...
~RSUMMARY: AddressSanitizer: SEGV (...)R~
~R         in AcseConnection_parseMessageR~
\end{lstlisting}
        \end{minipage}
    \end{lrbox}
    \begin{lrbox}{\physGuardPanel}
        \begin{minipage}[t]{0.48\linewidth}
        \vspace{0pt}
        \codehdr{iso\_acse/acse.c \hfill AcseConnection\_parseMessage}
\begin{lstlisting}[style=dvDiff]
 uint8_t* buffer = message->buffer;
 int messageSize = message->size;
 int bufPos = 0;
+if ((buffer == NULL) || (messageSize < 1))
+{
+    /* ... DEBUG_ACSE printf omitted ... */
+    return ACSE_ERROR;
+}
 ~Ruint8_t messageType = buffer[bufPos++];R~
\end{lstlisting}
        \end{minipage}
    \end{lrbox}
    \begin{lrbox}{\physReviewPanel}
        \begin{minipage}[t]{0.48\linewidth}
        \vspace{0pt}
        \codehdr{Patch Review Agent (summary)}
        \textpanel{%
            \dvitem{Verdict}{\texttt{FAIL\_VALIDATION\_SYMPTOM}.}
            \dvitem{Initial patch}{Prevents the NULL read but leaves the upstream success condition unchanged.}
            \dvitem{Upstream parser}{\texttt{IsoPresentation\_parseConnect} reports success without ACSE user data.}
            \dvitem{Required check}{Before returning success, require \texttt{nextPayload.buffer != NULL} and \texttt{nextPayload.size > 0}.}}
        \end{minipage}
    \end{lrbox}
    \begin{lrbox}{\physFixPanel}
        \begin{minipage}[t]{0.48\linewidth}
        \vspace{0pt}
        \codehdr{iso\_presentation.c \hfill IsoPresentation\_parseConnect}
\begin{lstlisting}[style=dvDiff]
 if (hasNormalModeParameters == false) {
     /* ... DEBUG_PRES printf omitted ... */
     return 0;
 }
+if ((self->nextPayload.buffer == NULL) ||
+    (self->nextPayload.size < 1)) {
+    /* ... DEBUG_PRES printf omitted ... */
+    return 0;
+}
 return 1;
\end{lstlisting}
        \end{minipage}
    \end{lrbox}
    \setlength{\physTopRowHeight}{\dimexpr\ht\physReportPanel+\dp\physReportPanel\relax}
    \ifdim\dimexpr\ht\physGuardPanel+\dp\physGuardPanel\relax>\physTopRowHeight
        \setlength{\physTopRowHeight}{\dimexpr\ht\physGuardPanel+\dp\physGuardPanel\relax}
    \fi
    \setlength{\physBottomRowHeight}{\dimexpr\ht\physReviewPanel+\dp\physReviewPanel\relax}
    \ifdim\dimexpr\ht\physFixPanel+\dp\physFixPanel\relax>\physBottomRowHeight
        \setlength{\physBottomRowHeight}{\dimexpr\ht\physFixPanel+\dp\physFixPanel\relax}
    \fi

    \begin{subfigure}[t]{0.49\linewidth}
        \begin{minipage}[t][\physTopRowHeight][t]{\linewidth}
            \vspace{0pt}
            \noindent\usebox{\physReportPanel}
        \end{minipage}
        \captionsetup{margin=0.3em}
        \caption{AddressSanitizer report for CVE-2021-45769 (NULL-pointer read).}
        \label{subfig:san_45769}
    \end{subfigure}
    \hfill
    \begin{subfigure}[t]{0.49\linewidth}
        \begin{minipage}[t][\physTopRowHeight][t]{\linewidth}
            \vspace{0pt}
            \noindent\usebox{\physGuardPanel}
        \end{minipage}
        \captionsetup{margin=0.3em}
        \caption{Crash-site guard: passes the PoC and code-based functionality assessment.}
        \label{subfig:guard_45769}
    \end{subfigure}

    \vspace{0.6em}

    \begin{subfigure}[t]{0.49\linewidth}
        \begin{minipage}[t][\physBottomRowHeight][t]{\linewidth}
            \vspace{0pt}
            \noindent\usebox{\physReviewPanel}
        \end{minipage}
        \captionsetup{margin=0.3em}
        \caption{Review rejects (b) and targets the upstream parser's success condition.}
        \label{subfig:review_45769}
    \end{subfigure}
    \hfill
    \begin{subfigure}[t]{0.49\linewidth}
        \begin{minipage}[t][\physBottomRowHeight][t]{\linewidth}
            \vspace{0pt}
            \noindent\usebox{\physFixPanel}
        \end{minipage}
        \captionsetup{margin=0.3em}
        \caption{Upstream check: rejects missing payloads and passes the PoC and final review.}
        \label{subfig:fix_45769}
    \end{subfigure}

    \caption{Physically distant vulnerability in libiec61850 (CVE-2021-45769): review redirects a PoC-passing repair to an upstream parser.}
    \label{Fig:motivation_phys}
\end{figure}

\subsection{Physically Distant Defect (CVE-2021-45769)}
\label{ss:physical}
Physical distance arises when the root cause lies outside the crash-site function.
An earlier operation may return before its invalid output causes a crash, so identifying the root cause requires tracing beyond the active stack.

\PP{Crash site and upstream cause.}
In CVE-2021-45769, a NULL-pointer dereference occurs in libiec61850's \texttt{AcseConnection\_parseMessage} (\autoref{subfig:san_45769}).
The caller passes the input buffer after \texttt{IsoPresentation\_parseConnect} reports success.
However, that parser can accept a connect PDU without ACSE user data, leaving \texttt{nextPayload} empty with a NULL buffer.
The presentation parser has already returned when the crash occurs and is therefore absent from the crash stack.

\PP{A PoC-passing candidate.}
\sysname{} first guards the faulting read, returning \texttt{ACSE\_ERROR} for a NULL or empty buffer (\autoref{subfig:guard_45769}).
The candidate passes the PoC and the Functionality Validation Agent's code-based assessment; no functionality test script is available.
Yet the upstream parser still reports success without supplying the required payload.

\PP{Repair guided by patch review.}
The Patch Review Agent rejects the patch for leaving this condition unresolved and directs repair to \texttt{IsoPresentation\_parseConnect} (\autoref{subfig:review_45769}), which initial localization had already identified as a candidate.
The next patch requires a non-NULL, non-empty payload before returning success (\autoref{subfig:fix_45769}).
It passes the PoC, code-based functionality assessment, and final review, completing repair after two candidates.
Review thus prompts revision of a patch that an acceptance policy based only on the preceding checks would have accepted.

\subsection{Summary}
The two examples motivate complementary mechanisms in \sysname{}.
Patch-conditioned re-localization uses an intermediate failure to investigate a logically distant defect, while root-cause-oriented patch review identifies an unresolved upstream cause in a PoC-passing patch.
Together, these mechanisms go beyond pass/fail validation by using changed execution failures to guide re-localization and reviewing test-passing patches for unresolved root causes.
\section{Design}
\label{s:Design}

\begin{figure}[t]
  \centering
  \includegraphics[width=0.6\linewidth]{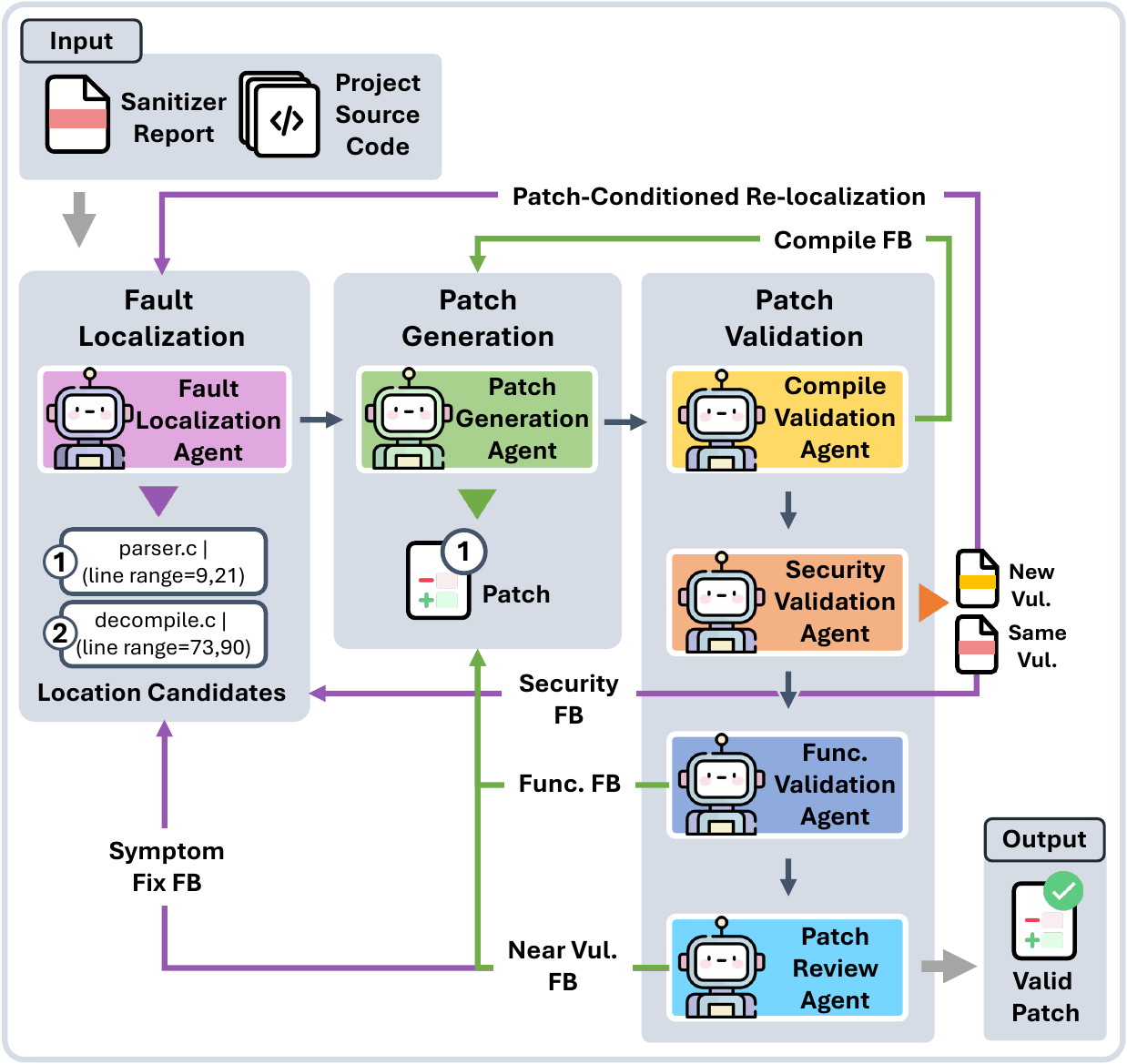}
    \caption{Overall Workflow of \sysname. The abbreviation FB stands for Feedback.}
  \label{fig:overview}
\end{figure}

\PP{Problem setup.}
 We focus on patching security vulnerabilities in realistic settings where the repair system operates on the full software codebase and a single PoC test case (i.e., a reliably reproducible bug trigger) that exposes the vulnerability. 
We note that not all automated program repair approaches adhere to this setting, as many assume access to exact fault localization ~\cite{APR_icse26_Flames, APR_icse26_Reinfix, APR_ISSTA24_ChatRepair, APR_ISSTA24_thinkrepair, AVR_ASE25_MUSSEL, AVR_ASE25_Vul-R2, icse24, fse22}, user-provided specifications, or explicit repair constraints.

In this section, we describe the design of \sysname{}. We first present the overall multi-agent framework, each agent in the framework, and their feedback mechanisms (\autoref{ss:agent_roles}). We then introduce patch-conditioned re-localization (\autoref{ss:patch-cond}) and the Patch Review Agent (\autoref{ss:Patch_Review}), which together enable \sysname{} to perform deep analysis across iterations, thereby producing effective patches even when the initial sanitizer report offers only weak signals due to physical or logical distance.

\PP{Overview.}
\autoref{fig:overview} depicts the end-to-end workflow of \sysname{}. Following the standard AVR decomposition into localization, patch generation, and validation~\cite{SoKavr,li2025sokeffectiveautomatedvulnerability}, \sysname{} realizes these stages via six role-specific LLM agents that exchange structured feedback.
(1) The process begins with a sanitizer report and the project's source code. The \emph{Fault Localization Agent} analyzes these artifacts to identify and rank code regions that are likely to contain the root cause. (2) Given a selected candidate location and a root-cause hypothesis, the \emph{Patch Generation Agent} produces a candidate patch. (3) The \emph{Compile Validation Agent} applies the patch to the codebase and checks build success. Compilation failures are analyzed and returned as feedback for refinement. (4) Once the program builds, the \emph{Security Validation Agent} reruns the proof-of-concept (PoC) to determine whether the vulnerability is eliminated. If the PoC still triggers the same vulnerability type, the agent summarizes the failure evidence and extracts actionable feedback, then returns it to the Fault Localization Agent to revise the location or strategy. If a different vulnerability type is triggered, \sysname{} keeps the current patch applied and re-invokes the \emph{Fault Localization Agent} on the updated sanitizer report (patch-conditioned re-localization, \autoref{ss:patch-cond}). (5) If the generated patch passes the security validation, the \emph{Functionality Validation Agent} runs regression tests to check that the patch preserves intended behavior. When tests fail, the agent analyzes the failing cases to diagnose the likely cause of the regression and returns targeted feedback to the Patch Generation Agent for patch refinement. (6) Finally, the \emph{Patch Review Agent} (\autoref{ss:Patch_Review}) performs code-review style checks to ensure the patch is root-cause oriented and to surface nearby vulnerability risks.
This feedback-driven loop continues until \sysname{} finds a valid patch or reaches a predefined iteration limit (15 in our experiments).

\subsection{Failure-Aware Role Decomposition and Feedback Routing}
\label{ss:agent_roles}
\sysname{}'s first design principle is failure-aware role decomposition. Rather than treating intermediate repair outcomes as a binary validation signal, \sysname{} distinguishes compilation, security, functionality, and review failures, routing each as diagnostic feedback to the stage best suited to correct the error. This subsection details how responsibilities, tool access, and feedback outputs are assigned across the six agents, enabling the system to recover evidence even when the root cause is distant from the crash site.

\PP{Role Decomposition.}
Each agent is controlled by a dedicated system prompt with a shared rule-based format specifying its objective, permitted tools, decision rules, and output schema. This keeps agents aligned with their roles, enforces least-privilege tool use, and standardizes the feedback exchanged across stages. Condensed versions of each agent's prompt are provided in \autoref{app:prompts}.

\textit{Fault Localization Agent}
moves beyond the crash site by analyzing the sanitizer report and tracing crash-related objects through the call stack. To recover evidence under weak signals, it inspects entry-to-call-site regions, retrieves full function context, and performs bounded expansion into sibling callees when earlier side effects may influence the failure. It returns a ranked list of candidate locations with concise rationales based on data-flow dependencies, violated invariants, and allocation-to-use mismatches.

\textit{Patch Generation Agent}
converts a selected candidate location and root-cause hypothesis into a minimal, scope-valid edit. It prioritizes fixes at the origin of the vulnerability, allowing guard-only or early-exit patches only as a fallback with explicit justification. To reduce malformed edits and hallucinated context, the agent re-reads the exact patch region immediately before emission and enforces structural validity, including balanced scopes and consistent control-flow structure. 

\textit{Compile Validation Agent}
is restricted to patch application and recompilation. It cannot modify code beyond the provided patch or create new files. On failure, it summarizes the actionable diagnostics and maps them to likely structural issues in the patch, such as malformed replacement, missing declarations, or type mismatches, enabling targeted revision by the patch generation agent.

\textit{Security Validation Agent} evaluates the patched program solely from a security perspective by rerunning the PoC under sanitizer instrumentation. It reports whether the sanitizer error disappears, persists with the same vulnerability type, or changes to a different type, and returns a concise evidence summary plus bug-type analysis to guide downstream routing. In particular, a changed vulnerability type triggers patch-conditioned re-localization.

\textit{Functionality Validation Agent}
checks behavioral preservation by running regression tests when available. Instead of returning raw logs, it diagnoses likely causes of behavioral regressions, such as overly restrictive checks, altered control flow, or changed assumptions, and returns targeted guidance to patch generation. When tests are unavailable, it performs a lightweight patch-impact analysis based on the surrounding code.

\textit{Patch Review Agent}
performs code-review-style checks beyond PoC execution. It flags symptom-oriented edits, detects misplaced validation logic, and checks patch completeness across nearby, logically equivalent sites. It returns either a pass decision or concrete revision suggestions grounded in the surrounding code context.

\begin{table}[t!]
\centering
\caption{Tools invoked by \sysname.}
\label{tab:tools}
\footnotesize

\scriptsize
\begin{tabular}{p{2.5cm}p{8.5cm}}
\hline
\textbf{Tool} & \textbf{Description} \\
\hline
\multicolumn{2}{l}{\textbf{Context Extraction}} \\
\hline
\texttt{FileReadLines} & \makecell[l]{Reads up to 100 lines from a specified file to provide local code context.} \\
\texttt{GetFunction} & \makecell[l]{Given a line number, returns the enclosing function's name, range, and content.} \\
\texttt{\makecell[l]{GetEntrytoCallSite}} & \makecell[l]{Extracts code from the function's entry point to a given call site line.} \\
\texttt{GetDirectory} & \makecell[l]{Returns files and directories in the specific path.} \\
\texttt{\makecell[l]{Repository\_retriever}} & \makecell[l]{Retrieves semantically relevant code blocks.} \\
\hline
\multicolumn{2}{l}{\textbf{Execution}} \\
\hline
\texttt{PatchCompile} & \makecell[l]{Resets the repository, applies a patch, and recompiles the project.} \\
\texttt{PoCExecution} & \makecell[l]{Executes the binary with the PoC to validate security.} \\
\texttt{\makecell[l]{FuncTestExecution}} & \makecell[l]{Executes regression tests to validate functionality preservation.} \\
\hline
\end{tabular}
\end{table}

\PP{Tool Support.}
Table~\ref{tab:tools} summarizes the tools available to \sysname{}. They fall into two groups: context-extraction tools for repository navigation and code retrieval, and execution tools for controlled patch application, recompilation, PoC execution, and functionality test execution. 

Agents retrieve and expand relevant code context through tool calls during reasoning, rather than relying solely on their initial inputs. These tools are part of the analysis process, not just execution utilities: in particular, the Fault Localization Agent uses entry-to-call-site retrieval, full-function lookup, and repository search to follow crash-related objects beyond the immediate stack trace under weak sanitizer signals.

Access is role-specific. Fault localization, patch generation, and patch review agents are restricted to read-only context-extraction tools, while the compile, security, and functionality validation agents are limited to their designated execution tools and cannot arbitrarily modify code. This least-privilege design reduces cross-stage interference and keeps each agent focused on its assigned role.

\begin{figure}[t!]
    \centering
    \includegraphics[width=0.6\linewidth]{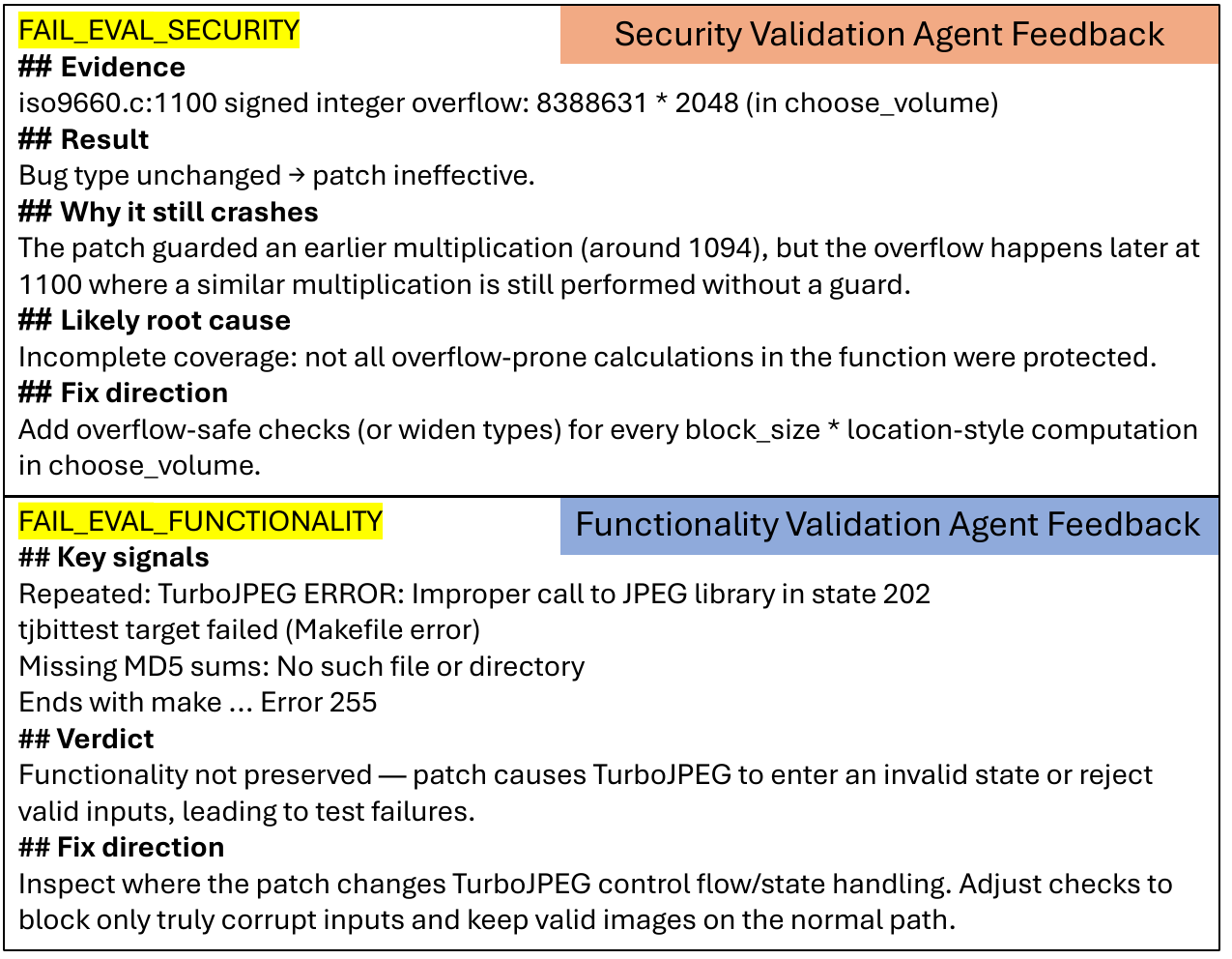}
    \caption{Security and functionality validation feedback produced by \sysname{}’s validation agents.}
    \label{fig:feedback}
\end{figure}

\PP{Agent Feedback.}
Beyond role specialization, \sysname{} treats intermediate outcomes, including failures, as first-class diagnostic signals. After each patch attempt, four specialized validation agents interpret the execution results, extract the most informative evidence, and route actionable feedback to an upstream agent. Each feedback message begins with a routing keyword (e.g., \texttt{PASS\_EVAL\_SECURITY}, \texttt{FAIL\_EVAL\_SECURITY}), which is mapped to a predefined transition in the agent graph. \autoref{fig:feedback} shows representative feedback messages produced by the Security and Functionality Validation Agents, where the keyword header is followed by an evidence excerpt and a concise diagnosis with a suggested fix direction. Concretely, compilation and regression failures are routed to the Patch Generation Agent because they indicate malformed edits or semantic regressions, while security failures are routed to the Fault Localization Agent because they suggest mislocalization or an ineffective repair strategy. Patch Review Agent routes its feedback to either the Patch Generation Agent (to revise the edit) or the Fault Localization Agent (to re-localize), depending on whether it detects a patch-quality issue or evidence of mislocalization.

\subsection{Patch-Conditioned Re-localization}
\label{ss:patch-cond}
\sysname{} incorporates patch-conditioned re-localization to leverage sanitizer feedback even when an intermediate patch changes the observed failure. This is important because agent-generated patches can (i) fail to eliminate the original vulnerability or (ii) introduce another vulnerability whose type could even be different, yielding a new sanitizer signature that reveals additional context about the program state and execution path.

Algorithm~\autoref{alg:pcrl} summarizes the procedure. After a candidate patch $P_1$ passes compilation, the Security Validation Agent inspects the resulting sanitizer report $R_1$. If the same vulnerability type reappears, \sysname{} treats the patch as ineffective and feeds a concise description of failure back to the Fault Localization Agent, which then expands or revises the candidate set, for example, by exploring alternative locations or revisiting a previous location with a different repair strategy. In contrast, if a different vulnerability type emerges (line~10), \sysname{} first asks the Functionality Validation Agent whether $P_1$ preserves the intended behavior (line~11), so that re-localization does not build on a patch that already breaks the program. It then restarts the workflow on the repository state with $P_1$ applied (line~12). This allows the Fault Localization Agent to analyze $R_1$ in the updated context and localize the defect responsible for the newly observed failure. By re-localizing under the patched program state, \sysname{} progressively shifts attention from symptom sites toward deeper root-cause hypotheses, especially in cases where the observed failure is logically distant from its initiating defect. If re-localization fails to produce a validated patch $P_2$, \sysname{} takes this as evidence that $P_1$ sits at the wrong site and returns to fault localization on the original program (line~13).

\providecommand{\IfThen}[2]{\State \algorithmicif\ #1\ \algorithmicthen\ #2}
\begin{algorithm}[t]
\caption{Patch-conditioned re-localization}
\label{alg:pcrl}
\footnotesize
\begin{algorithmic}[1]
\Require Program $S$, PoC input $x$, sanitizer report $R_0$ of $S$ on $x$
\Ensure Final patch $P$, or \textsc{Fail}
\Function{Repair}{$S, R$}
  \Loop \Comment{bounded by the iteration limit}
    \State $P_1 \gets$ \Call{LocalizeAndPatch}{$S, R$}
    \IfThen{\Call{Compile}{$S \oplus P_1$} fails}{\textbf{continue}}
    \State $R_1 \gets$ \Call{RunPoC}{$S \oplus P_1, x$}
    \If{$R_1$ is empty} \Comment{crash removed}
      \IfThen{\Call{FuncAgent}{$P_1$} $\wedge$ \Call{ReviewAgent}{$P_1$}}{\Return $P_1$}
      \State \textbf{continue}
    \EndIf
    \IfThen{\Call{SecAgent}{$R, R_1$} $\neq$ \textsc{BugTypeChanged}}{\textbf{continue}}
    \IfThen{$\neg$\Call{FuncAgent}{$P_1$}}{\textbf{continue}}
    \State $P_2 \gets$ \Call{Repair}{$S \oplus P_1, R_1$} \Comment{re-localize on the patched program}
    \IfThen{$P_2 = \textsc{Fail}$}{\textbf{continue}} \Comment{$P_1$ is likely at the wrong site}
    \State $d \gets$ \Call{Select}{$S, P_1, P_2, x$}
    \If{$d = \textsc{Drop}$} \Return $P_2$
    \ElsIf{$d = \textsc{Keep}$} \Return $P_1 \oplus P_2$
    \EndIf
    \Comment{$d = \textsc{Symptom}$: re-localize with the review feedback}
  \EndLoop
  \State \Return \textsc{Fail}
\EndFunction
\Statex
\Function{Select}{$S, P_1, P_2, x$}
  \State $v \gets \textsc{False}$ \Comment{is the re-localized patch sufficient on its own?}
  \If{$P_1$ can be reverted from $S \oplus P_1 \oplus P_2$}
    \State $S' \gets S \oplus P_2$
    \State $v \gets$ \Call{Compile}{$S'$} $\wedge$ \Call{RunPoC}{$S', x$} is empty $\wedge$ \Call{FuncAgent}{$P_2$}
  \EndIf
  \State $d \gets$ \Call{ReviewAgent}{$P_1, P_2, v$} \Comment{check 4 in Fig.~\ref{fig:prompt_PR}}
  \IfThen{$d = \textsc{Drop} \wedge \neg v$}{$d \gets \textsc{Keep}$} \Comment{execution overrides the agent}
  \State \Return $d$
\EndFunction
\end{algorithmic}
\end{algorithm}

When re-localization succeeds, the resulting program contains both the intermediate patch $P_1$ and the re-localized patch $P_2$. However, $P_1$ was generated without knowledge of the deeper defect, so it may play one of two roles: it may be a necessary part of the fix, for example when the developer patch also modifies both locations, or a diagnostic probe that merely exposed the deeper failure and became redundant once $P_2$ fixed it. Shipping both patches unconditionally would retain such probes, which are often symptom-level guards. \sysname{} therefore selects the final patch through execution rather than by the judgment of an agent alone (\textsc{Select}, lines~21--30). It reverts $P_1$ from the patched program, rebuilds it, reruns the PoC, and invokes the Functionality Validation Agent on the remaining patch (lines~23--25). The outcome $v$ indicates whether $P_2$ alone is a sufficient fix. The Patch Review Agent then decides whether to keep $P_1$ (\autoref{ss:Patch_Review}), but the workflow lets it drop $P_1$ only when $v$ holds (line~28).

\definecolor{prBorder}{HTML}{8CC8E8}
\begin{figure}[t]
\centering
\setlength{\fboxsep}{5pt}
\setlength{\fboxrule}{1pt}
\fcolorbox{prBorder}{white}{%
\begin{minipage}{0.85\linewidth}
\footnotesize\sffamily\raggedright
Determine whether a patch reflects a \emph{developer-level fix} by checking:\\[1pt]
\textbf{1) Early-Exit Detection (Symptom Fix)}\\
Inspect newly added guards or early exits.\\
If a guard merely avoids the crashing path without enforcing the true invariant,
restoring state, or propagating errors consistently, classify it as a \emph{symptom-level fix}.\\[1pt]
\textbf{2) Right-Layer Validation (Symptom Fix)}\\
Identify where the guarded value is created or becomes invalid.\\
Check if validation and error handling are at that source layer
(caller/producers/parsers), not solely at the sink where the sanitizer reports the crash.\\[1pt]
\textbf{3) Near-Vulnerability Completeness}\\
Analyze the full function and identify all code paths sharing the same invariant and risky operation.\\
Fixes applied to only a subset of equivalent sites are considered incomplete.\\[1pt]
\textbf{4) Patch Selection (after patch-conditioned re-localization)}\\
Decide whether the intermediate patch $P_1$ is still needed alongside the re-localized patch $P_2$.\\
If $P_2$ alone passes validation, drop $P_1$ as a diagnostic probe;
otherwise keep $P_1$ unless checks 1 or 2 classify it as a \emph{symptom-level fix}.
\end{minipage}}
\caption{Tailored prompt for the Patch Review Agent.}
\label{fig:prompt_PR}
\end{figure}

\subsection{Root-cause-oriented Patch Review}\label{ss:Patch_Review}
Execution-based validation alone cannot guarantee patch correctness: passing the PoC and available regression tests only confirms that the observed failure is no longer triggered, not that the underlying vulnerability has been fully eliminated. This limitation is especially severe for physically distant vulnerabilities, where a local guard at the crash site may silence the sanitizer report while upstream invalid state continues to propagate. To address this gap, \sysname{} includes a \emph{root-cause-oriented Patch Review Agent} that screens validated patches for symptom-level and incomplete repairs before they are accepted.

Given a candidate patch that already compiles and passes security and functionality validation, the Patch Review Agent performs a code-centric audit over the full patched function. It focuses on three checks, summarized in \autoref{fig:prompt_PR}: \textbf{(i)} \emph{Early-exit Detection}, which flags early-exit edits that avoid the crash without enforcing the true safety invariant; \textbf{(ii)} \emph{Right-Layer Validation}, which checks whether the patch prevents the invalid state at its source (e.g., caller/parser/producer), rather than adding guards only at the sanitizer-reported sink (the use site); and \textbf{(iii)} \emph{Near-Vulnerability Completeness}, which detects logically equivalent risky sites in the same function that remain unfixed. This check is motivated by developer patches that generalize a repair across adjacent, structurally similar branches once the shared invariant is identified; for example, in \autoref{subfig:dev_8964}, the developer fixes both \texttt{PUSH\_CONSTANT} and \texttt{PUSH\_CONSTANT16}, even though the PoC triggers only one branch, because both share the same bounds invariant.

Based on these checks, the agent returns either \texttt{PASS\_VALIDATION} or revision feedback for the next iteration. When the fix is an early-exit patch or validated at the wrong layer, the feedback is routed to the Fault Localization Agent to revisit the repair site. When the patch is correct but incomplete, the feedback is routed to the Patch Generation Agent to propagate the fix across equivalent nearby sites. In this way, patch review acts as a final diagnostic layer that pushes repairs beyond crash suppression toward root-cause-correct patches.

\PP{Patch Selection.}
After a successful patch-conditioned re-localization, the re-localized patch $P_2$ has already passed the three checks above. The Patch Review Agent is then invoked once more, with a dedicated prompt (check 4 in \autoref{fig:prompt_PR}), to decide whether the intermediate patch $P_1$ belongs in the final patch (\textsc{Select} in Algorithm~\autoref{alg:pcrl}). Its input consists of both patches, the original and intermediate sanitizer reports, and the execution result of $P_2$ alone. If $P_2$ alone passes, the agent drops $P_1$ as a diagnostic probe, unless $P_1$ fixes a concrete defect that $P_2$ does not. Otherwise, that is, if $P_2$ alone fails or $P_1$ cannot be reverted because both patches edit the same lines, $P_1$ is required, and the agent applies checks 1 and 2 to it: if $P_1$ is not a symptom-level fix, the agent keeps it and \sysname{} returns $P_1 \oplus P_2$; if it is, the feedback is routed to the Fault Localization Agent (line~17 of Algorithm~\autoref{alg:pcrl}), because the original failure is still only suppressed.

\section{Experimental Setup}
\label{s:Experimental_setup}

We evaluate the vulnerability repair capability of \sysname{} by answering the following key research questions:

\begin{itemize}[leftmargin=*,label={}]
    \item \textbf{RQ1)} How effectively can \sysname{} repair vulnerabilities in real-world programs? (\autoref{ss:rq1})
    \item \textbf{RQ2)} How well does \sysname{} generalize to vulnerabilities unseen during training? (\autoref{ss:rq2})
    \item \textbf{RQ3)} How effectively can \sysname{} repair vulnerabilities whose fix lies away from the crash site? (\autoref{ss:rq3})
    \item \textbf{RQ4)} How does each specialized agent of \sysname{} contribute to the overall performance of the AVR task? (\autoref{ss:rq4})
    \item \textbf{RQ5)} What are the computational cost and runtime overhead of \sysname{} in practice? (\autoref{ss:rq5})
\end{itemize}

\subsection{Benchmark}
\label{ss:benchmark}

We first evaluate \sysname{} on the VulnLoc benchmark~\cite{shen2021localizing}, following prior end-to-end AVR studies~\cite{kim2025logs,looprepair,crashrepair} to enable direct comparison. Following SAN2PATCH~\cite{kim2025logs}, we use 39 of the 43 VulnLoc samples, excluding four samples that cannot be reproduced with sanitizer instrumentation. The evaluated vulnerabilities span six bug classes: Integer-Overflow (IO), Buffer-Overflow (BO), Divide-by-Zero (DZ), Null-Pointer-Dereference (NPD), Data-Type-Overflow (DT), and Use-After-Free (UAF).

To evaluate generalization (RQ2, see~\autoref{ss:rq2}), we use the SAN2VULN benchmark~\cite{kim2025logs}, which contains 27 vulnerabilities from four open-source projects, including several recently discovered cases for which no developer patch is yet available. Compared with the evaluated VulnLoc samples, SAN2VULN includes an additional bug class: Memory-Leak (ML). For our experiments, we filtered the benchmark by fix commit date, retaining only vulnerabilities with fixes introduced after June 2024 to mitigate potential training data leakage in \texttt{gpt-4.1}. This excluded two \texttt{liblouis} cases, leaving 25 samples.

To evaluate \sysname{}'s ability to repair vulnerabilities whose fix lies away from the crash site (RQ3, see~\autoref{ss:rq3}), we use the 200-CVE evaluation split of SEC-bench~\cite{secbench}, a benchmark designed to assess LLM agents on real-world software security tasks, including vulnerability repair. Of these 200 instances, 112 (56\%) have a developer fix that does not modify the crash-site function. We retain SAN2PATCH and PatchAgent from the VulnLoc evaluation and additionally include DebugHarness~\cite{debugharness} and EvoRepair~\cite{evorepair}, two recent agent-based vulnerability repair methods previously evaluated on SEC-bench. 
Here, we use GPT-5-mini~\cite{openai2025gpt5mini} to align with prior SEC-bench evaluations~\cite{evorepair} and DeepSeek-V4.1-Flash~\cite{deepseekai2026v41flash} because its lower API cost makes re-executing methods on all 200 instances feasible. 
We evaluate each method with both backends on the same instance set, except for EvoRepair's GPT-5-mini result, for which we use the aggregate reported in the original paper.

\subsection{Baselines and Evaluation Metrics}
\paragraph{Baselines.}
For the evaluation on VulnLoc, we compare \sysname{} against the following five state-of-the-art methods for end-to-end vulnerability repair, which are open-source systems that operate on C/C++ codebases and require only a single failing test input, namely a PoC or its corresponding bug report (e.g., sanitizer log).

\begin{itemize}[leftmargin=*, itemsep=0pt, topsep=0pt]
\item CrashRepair~\cite{crashrepair}. A program-analysis-based AVR system that leverages concolic execution to localize candidate fix locations and infer crash-free constraints from a crashing input. It then synthesizes patches by searching for code mutations that satisfy the translated constraints.

\item LoopRepair~\cite{looprepair}. An iterative AVR pipeline that combines vulnerability localization, LLM-based patch-hunk location prediction, location-aware patch generation, and taint-trace-guided patch ranking. It uses CrashAnalysis~\cite{crashrepair} for localization and iteratively selects the best failing patch for refinement based on taint-trace feedback.

\item SAN2PATCH~\cite{kim2025logs}. 
An LLM-based repair pipeline that decomposes vulnerability repair into staged reasoning over sanitizer logs using prompting strategies such as CoT~\cite{neurips22-cot}, SC~\cite{wang2022self}, and ToT~\cite{yao2023tree}. The framework iteratively retries patch generation when validation on the PoC or functional test suite fails.

\item PatchAgent~\cite{yu2025patchagent}. A single-agent AVR that uses tool APIs to iteratively retrieve code context, locate relevant definitions, and validate candidate patches. Starting from a sanitizer log, it leverages counterexample feedback to refine patches.

\item SpecRover~\cite{specrover}. A multi-agent APR that performs repository-level context retrieval, infers function-level specifications, and uses reviewer-guided feedback for iterative repair.

\end{itemize}
Although SpecRover is originally designed for GitHub issue resolution, it supports repair from sanitizer logs and has been shown to handle a sample from DARPA AI Cyber Challenge (AIxCC). To adapt SpecRover for AVR, we disable its reproducer agent and directly use the provided PoC input, since each vulnerability already comes with an executable crashing test case, and modify its retrieval/parsing pipeline to support C projects, as detailed in \autoref{sec:app_specrover}.

For SAN2VULN, we restrict the comparison to SAN2PATCH, which achieved the highest correct-patch count among the baselines on VulnLoc, as detailed in \autoref{ss:rq2}. For SEC-bench, we compare \sysname{} against SAN2PATCH, PatchAgent, DebugHarness~\cite{debugharness}, and EvoRepair~\cite{evorepair}, as detailed in \autoref{ss:rq3}.

\paragraph{Evaluation Metrics.}
For VulnLoc and SAN2VULN, we use two commonly adopted metrics following prior work~\cite{ye2024iter,looprepair,crashrepair,kim2025logs}: the number of plausible patches and the number of correct patches. A patch is considered \emph{plausible} ($\orangetriangle$) if it passes the PoC and, when available, the full functional test suite. However, passing these tests does not necessarily guarantee security or functional correctness~\cite{sp23, han2025rethinking}. We therefore include manual validation of all plausible patches, assessing correctness against developer patches where available and against root-cause repair criteria otherwise. When a developer patch is available, a plausible patch is deemed \emph{correct} (\checkmark) if it is semantically equivalent to the developer’s patch. We further classify plausible but incorrect patches into two categories: Not Secure ($\orangetriangle$-NS), where the patch suppresses the PoC without addressing the root cause, and Not Functionally Correct ($\orangetriangle$-NF), where the patch breaks functionality under valid inputs.

To ensure consistency in manual evaluation, we apply the following correctness criteria:
(1)~\textit{Root-cause alignment}: the patch addresses the earliest point within the execution path leading to the crash where the invariant is first violated, regardless of which function it resides in, rather than suppressing symptoms at the crash site.
(2)~\textit{Functional preservation}: the patch does not alter the intended behavior of the program for valid inputs.
For samples with a developer patch, correctness is verified by semantic equivalence to the ground truth. For samples without a developer patch (some samples in SAN2VULN), correctness is determined solely by the above criteria.

For SEC-bench, we report benchmark acceptance under the Medium criterion, which requires the patched program to run the PoC without sanitizer errors and exit either with code 0 or with the same code as
the reference-patched program.
We additionally report file- and function-level localization coverage, as detailed in \autoref{ss:rq3}.

\subsection{Implementation}
\sysname{} is implemented in Python~3.10 using LangChain~\cite{langchain}. Each agent is instantiated as an independent LangChain agent, and the overall workflow is represented as a directed graph: nodes correspond to agents and edges encode allowable hand-offs between agents. Importantly, the routing policy is implemented in code (graph edges + routing functions). 
Each agent is instructed (via its system prompt) to start its response with exactly one \emph{routing keyword} from a fixed set (e.g., \texttt{PASS\_EVAL\_SECURITY}). At runtime, the controller parses this keyword and deterministically dispatches execution to the next agent according to a predefined mapping.

The \texttt{repository\_retriever} tool follows the approach of RLCoder, a repository-level code completion tool~\cite{wang2025rlcoder}. It is built with PyTorch~2.6.0~\cite{pytorch} and retrieves semantically relevant code blocks using UniXcoder~\cite{unixcoder} embeddings with FAISS-based indexing~\cite{faiss}. 
We ran our experiments on a 64-bit Ubuntu~20.04 system equipped with an Intel Xeon Silver~4316 CPU (20 hardware cores, 2.30\,GHz), 384\,GB of RAM, and four NVIDIA A6000 GPUs.

For VulnLoc and SAN2VULN, we used OpenAI's \texttt{gpt-4.1}~\cite{chatgpt} and re-executed the LLM-based comparison methods with the same backend. 
For SEC-bench, we evaluate methods with \texttt{GPT-5-mini}~\cite{openai2025gpt5mini}, following
EvoRepair~\cite{evorepair}, and \texttt{DeepSeek-V4.1-Flash}~\cite{deepseekai2026v41flash}, whose lower API
cost makes full evaluation on all 200 instances feasible.
\section{Experimental Results}
\label{s:Experimental_results}

\begin{table*}[ht]
\tiny
\centering
\caption{The number of plausible patches and correct patches generated by \sysname{} and the prior works on VulnLoc Benchmark~\cite{shen2021localizing}. PD and LD denote physically distant and logically distant cases, respectively.}
\label{tab:RQ1}
\newcolumntype{g}{>{\columncolor[gray]{0.95}}c}
\newcolumntype{h}[1]{>{\columncolor[gray]{0.95}\centering\arraybackslash}p{#1}}

\begin{tabular}{c c >{\centering\arraybackslash}p{1cm}>{\centering\arraybackslash}p{1cm} c c c c c g}
\hline
Package & Bug ID
& \makecell[c]{Distant\\Bug}
& \makecell[c]{Bug\\Type}
& \makecell[c]{CrashRepair\\\cite{crashrepair}}
& \makecell[c]{LoopRepair\\\cite{looprepair}}
& \makecell[c]{SAN2PATCH\\\cite{kim2025logs}}
& \makecell[c]{PatchAgent\\\cite{yu2025patchagent}}
& \makecell[c]{SpecRover\\\cite{specrover}}
& \cellcolor[gray]{0.95}\makecell[c]{\sysname\\(Ours)} \\
\hline

\multirow{4}{*}{binutils}
& CVE-2017-14745 & PD & IO  & \ding{55}            & \ding{55}            & \checkmark          & \ding{55}            & \checkmark          & \checkmark\\
& CVE-2017-15020 & -  & BO  & \ding{55}            & \ding{55}            & \checkmark          & \checkmark           & \checkmark          & \checkmark\\
& CVE-2017-15025 & -  & DZ  & $\orangetriangle$-NS & \ding{55}            & \checkmark          & \checkmark           & \ding{55}           & \checkmark \\
& CVE-2017-6965  & PD & BO  & $\orangetriangle$-NS & \ding{55}            & $\orangetriangle$-NF & \checkmark & \ding{55}           & \checkmark\\
\hline

\multirow{4}{*}{coreutils}
& gnubug-19784 & -  & BO & $\orangetriangle$-NF & \checkmark            & \checkmark          & \ding{55}            & \checkmark          & \checkmark\\
& gnubug-25003 & -  & IO & -            & -            & \checkmark          & $\orangetriangle$-NF & \checkmark          & \checkmark\\
& gnubug-25023 & PD & BO & $\orangetriangle$-NS & $\orangetriangle$-NS            & \checkmark          & \checkmark           & \checkmark          & \checkmark\\
& gnubug-26545 & -  & IO & -            & -            & $\orangetriangle$-NF & $\orangetriangle$-NF & $\orangetriangle$-NF & $\orangetriangle$-NF \\
\hline

\multirow{2}{*}{jasper}
& CVE-2016-8691 & PD & DZ & $\orangetriangle$-NF & $\orangetriangle$-NS & $\orangetriangle$-NS & $\orangetriangle$-NS & $\orangetriangle$-NS & \checkmark \\
& CVE-2016-9557 & -  & IO & $\orangetriangle$-NS & $\orangetriangle$-NS & \checkmark          & \checkmark           & \checkmark & \checkmark\\
\hline

libarchive
& CVE-2016-5844 & - & IO & $\orangetriangle$-NS & \ding{55}            & \checkmark          & \checkmark & \checkmark          & \checkmark \\
\hline

\multirow{4}{*}{libjpeg}
& CVE-2012-2806  & -  & BO  & \checkmark          & \checkmark          & \checkmark          & $\orangetriangle$-NS & \checkmark          & \checkmark \\
& CVE-2017-15232 & PD & NPD & \checkmark          & \ding{55}            & $\orangetriangle$-NF & $\orangetriangle$-NF & \checkmark          & \checkmark \\
& CVE-2018-14498 & -  & BO  & \ding{55}            & \ding{55}            & \checkmark          & $\orangetriangle$-NS & \checkmark          & \checkmark\\
& CVE-2018-19664 & PD & BO  & \ding{55}            & \ding{55}            & \ding{55}           & $\orangetriangle$-NF & $\orangetriangle$-NF & $\orangetriangle$-NF \\
\hline

\multirow{3}{*}{libming}
& CVE-2016-9264 & -  & BO  & $\orangetriangle$-NS & \ding{55}            & \checkmark          & \checkmark           & \ding{55}           & \checkmark \\
& CVE-2018-8806 & LD & UAF & \ding{55}            & \ding{55}            & \ding{55}           & \ding{55}            & \ding{55}           & \checkmark\\
& CVE-2018-8964 & LD & UAF & \ding{55}            & \ding{55}            & \ding{55}           & \ding{55}            & \ding{55}           & \checkmark\\
\hline

\multirow{13}{*}{libtiff}
& bugzilla-2611   & PD & DZ & \checkmark          & \checkmark          & \checkmark          & \checkmark          & \ding{55}           & \checkmark\\
& bugzilla-2633   & -  & BO & \checkmark          & $\orangetriangle$-NS & \checkmark          & $\orangetriangle$-NS & $\orangetriangle$-NS & \checkmark \\
& CVE-2016-10092  & PD & BO & \checkmark          & \checkmark          & \checkmark          & $\orangetriangle$-NS & \ding{55}           & \checkmark\\
& CVE-2016-10094  & PD & BO & \ding{55}            & $\orangetriangle$-NF & \checkmark          & $\orangetriangle$-NS & \checkmark          & \checkmark\\
& CVE-2016-10272  & PD & BO & \checkmark          & \ding{55}            & \checkmark          & $\orangetriangle$-NF & \ding{55}           & \checkmark\\
& CVE-2016-5321   & -  & BO & $\orangetriangle$-NF & \checkmark          & \checkmark          & $\orangetriangle$-NF & \ding{55}           & \checkmark\\
& CVE-2016-9273   & PD & BO & \ding{55}            & \ding{55}            & \ding{55}           & \checkmark          & \ding{55}           & \checkmark\\
& CVE-2016-9532   & LD & BO & $\orangetriangle$-NS & \ding{55}            & $\orangetriangle$-NS & $\orangetriangle$-NS & $\orangetriangle$-NS & $\orangetriangle$-NS \\
& CVE-2017-5225   & -  & BO & \checkmark          & \checkmark          & \checkmark          & $\orangetriangle$-NS & \ding{55}           & \checkmark\\
& CVE-2017-7595   & -  & DZ & $\orangetriangle$-NS & $\orangetriangle$-NS & \checkmark          & \checkmark          & \ding{55}           & \checkmark\\
& CVE-2017-7599   & -  & DT & \checkmark          & \checkmark          & \checkmark          & $\orangetriangle$-NS & \ding{55}           & \checkmark\\
& CVE-2017-7600   & -  & DT & \checkmark          & \checkmark          & \checkmark          & \checkmark          & \ding{55}           & \checkmark \\
& CVE-2017-7601   & -  & IO & $\orangetriangle$-NS & $\orangetriangle$-NS & \checkmark          & $\orangetriangle$-NS & \ding{55}           & \checkmark\\
\hline

\multirow{4}{*}{libxml2}
& CVE-2012-5134 & -  & BO  & \checkmark          & $\orangetriangle$-NS & \checkmark          & \checkmark          & \ding{55}           & \checkmark \\
& CVE-2016-1838 & -  & BO  & \checkmark          & \checkmark          & \checkmark          & $\orangetriangle$-NF & \ding{55}           & \checkmark\\
& CVE-2016-1839 & PD & BO  & \checkmark          & $\orangetriangle$-NS & \checkmark          & \checkmark          & \checkmark & \checkmark\\
& CVE-2017-5969 & -  & NPD & \checkmark          & \checkmark          & \checkmark          & \checkmark          & \ding{55}           & \checkmark \\
\hline

potrace
& CVE-2013-7437 & LD & BO & \checkmark & $\orangetriangle$-NS & $\orangetriangle$-NS & \checkmark & $\orangetriangle$-NS & \checkmark\\
\hline

\multirow{3}{*}{zziplib}
& CVE-2017-5974 & - & BO & \ding{55}            & $\orangetriangle$-NS            & \checkmark          & $\orangetriangle$-NS & $\orangetriangle$-NS & \checkmark\\
& CVE-2017-5975 & - & BO & \checkmark          & \ding{55}            & \checkmark          & \checkmark          & \checkmark          & \checkmark\\
& CVE-2017-5976 & - & BO & $\orangetriangle$-NS & \ding{55}            & \checkmark          & \checkmark          & $\orangetriangle$-NS & \checkmark\\
\hline

\multicolumn{4}{c}{Plausible Patches}
& 28/39 & 21/39 & 35/39 & 35/39 & 21/39 & \multicolumn{1}{g}{\textbf{39/39}} \\
\multicolumn{4}{c}{Correct Patches}
& 15/39 & 10/39 & 29/39 & 17/39 & 13/39 & \multicolumn{1}{g}{\textbf{36/39}} \\
\hline
\end{tabular}

\end{table*}

\subsection{Repair Success Rate on Real-world Vulnerability Benchmarks (RQ1)}
\label{ss:rq1}

\subsubsection{Candidate Budget.}
We set the maximum number of \sysname{}'s feedback loops to 15. If no valid patch is found, we repeat the run up to three times, and a vulnerability is counted as repaired only if a valid patch is produced within 45 candidates in total; this is the target budget for all baselines.
When patch-conditioned re-localization is triggered, re-localization runs its own feedback loop with its own loop limit; however, every candidate generated at any level counts toward the same per-vulnerability budget.
All methods are run until a valid patch is generated or the number of generated patches reaches 45.

SAN2PATCH generates up to nine candidates per run, so we execute it five times. PatchAgent is bound by tool invocations rather than patch count; in our reproduction on 39 VulnLoc samples, it produces about three patches per run on average, so we run it 15 times. SpecRover is configured with up to five internal iterations per round and at most 15 rounds.
For CrashRepair and LoopRepair, the number of generated patches depends on the number of fault-localization candidates passed to patch generation. CrashRepair runs until a valid patch is found or 45 candidates have been generated. LoopRepair, following the original setup, is capped at three iterations because the runtime grows substantially with additional iterations. If no valid patch is found within three iterations and fewer than 45 candidates have been generated, we run one additional execution to approach the same budget.

\subsubsection{LLM-based Repair Success Rate on VulnLoc}
\label{sss:vulnloc}

\autoref{tab:RQ1} presents the number of plausible and correct patches generated by \sysname{} and prior studies. For gnubug-25003 and gnubug-26545, initial localization was unavailable, so we leave CrashRepair and LoopRepair results empty.
The results show that \sysname{} consistently outperforms the other tools, generating the most plausible and correct patches in all bug types.
CrashRepair, LoopRepair, SAN2PATCH, PatchAgent, and SpecRover generated plausible patches for 28, 21, 35, 35, and 21 out of 39 cases, respectively, whereas \sysname{} generated plausible patches for all 39 cases.
In terms of correct patch generation, the performance gap widened further; CrashRepair, LoopRepair, SAN2PATCH, PatchAgent, and SpecRover produced 15, 10, 29, 17, and 13 valid repairs, respectively, whereas \sysname{} achieved 36 valid repairs (92.3\%).
These results demonstrate that \sysname{} not only generates more plausible patches but also correct fixes, addressing hidden flaws that prior systems often leave unresolved.
We note that none of the five comparison targets produced valid patches for CVE-2018-8806 and CVE-2018-8964, likely due to the logical distance between the initial sanitizer message and the root cause.
In contrast, \sysname{} successfully generates correct patches via deep analysis with patch-conditioned fault re-localization.

\begin{tcolorbox} [boxrule=0.8pt, top=0.2pt, bottom=0.2pt]
    \textbf{Answer to RQ1}: Compared to CrashRepair, LoopRepair, SAN2PATCH, PatchAgent, and SpecRover, \sysname{} achieves higher patch quality, producing plausible patches for 100\% of cases and correct patches for 92.3\%.
\end{tcolorbox}

\subsection{Repair Success Rate on Unseen Vulnerabilities (RQ2)} \label{ss:rq2}

\begin{table}[t!]
\tiny
\centering
\caption{Summary of unseen vulnerability repair results.}
\label{tab:table2}
\begin{tabular}{c c c c c}
\hline
Package
& Bug ID
& Bug Type
& \makecell[c]{SAN2PATCH\\[-1pt]\cite{kim2025logs}}
& \cellcolor[gray]{0.95}\makecell[c]{\sysname\\[-1pt](Ours)}
\\
\hline
\multirow{2}{*}{libming} & CVE-2024-24148 & ML & \checkmark & \checkmark \\
 & CVE-2024-24146 & ML & \checkmark & \checkmark \\
\hline
\multirow{9}{*}{Bento4} 
& NOFIX-2024-002 & BO & $\orangetriangle$-NS & $\orangetriangle$-NS \\
 & NOFIX-2024-003 & DZ & \checkmark & \checkmark \\
 & NOFIX-2024-004 & DZ & \checkmark & \checkmark \\
 & NOFIX-2024-005 & BO & \checkmark & \checkmark \\
 & NOFIX-2024-006 & BO & $\orangetriangle$-NS & \checkmark \\
 & NOFIX-2024-007 & DZ & \checkmark & \checkmark \\
 & NOFIX-2024-008 & NPD & $\orangetriangle$-NF & \checkmark \\
 & NOFIX-2024-009 & NPD & \checkmark & \checkmark \\
 & NOFIX-2024-010 & BO & \ding{55} & \checkmark \\
\hline
\multirow{10}{*}{Liblouis} 
& NOFIX-2024-001 & NPD & \checkmark & \checkmark \\
 & GIT-2024-1530 & NPD & $\orangetriangle$-NF & \checkmark \\
 & GIT-2024-1531 & NPD & \checkmark & \checkmark \\
 & GIT-2024-1532 & NPD & \checkmark & \checkmark \\
 & GIT-2024-1533 & NPD & $\orangetriangle$-NS & \checkmark \\
 & GIT-2024-1534 & BO  & $\orangetriangle$-NS & \checkmark \\
 & GIT-2024-1535 & BO  & $\orangetriangle$-NS & $\orangetriangle$-NS \\
 & GIT-2024-1536 & BO  & $\orangetriangle$-NS & $\orangetriangle$-NS \\
 & GIT-2024-1537 & BO  & \checkmark & \checkmark \\
 & GIT-2024-1539 & BO  & \checkmark & \checkmark \\
\hline
\multirow{4}{*}{Kamailio} & OSV-2024-1206 & BO & \checkmark & \checkmark \\
 & OSV-2024-1210 & BO & \checkmark & \checkmark \\
 & OSV-2024-1230 & BO & \checkmark & \checkmark \\
 & OSV-2024-1244 & BO & \checkmark & \checkmark \\
\hline
\multicolumn{3}{c}{Plausible Patches} & 24/25 & 25/25\\
\multicolumn{3}{c}{Correct Patches} & 16/25 & 22/25\\
\hline
\end{tabular}
\end{table}

To answer this research question, we use the SAN2VULN benchmark to evaluate \sysname{}’s ability to repair vulnerabilities that are unseen during training. Due to the high computational cost of running all four LLM-based baselines, we restrict this comparison to SAN2PATCH, which achieved the strongest performance in generating both plausible and correct patches in RQ1 (\autoref{ss:rq1}) among the comparison baselines. 
For unseen vulnerabilities, we report both plausible and correct patches, manually validating each patch to ensure it resolves the root cause rather than merely suppressing the PoC. As shown in \autoref{tab:table2}, \sysname{} produces 25 plausible patches, slightly exceeding SAN2PATCH’s 24. More importantly, \sysname{} yields fewer plausible but test-specific patches than SAN2PATCH (3 vs. 8 cases) under manual inspection.
These results indicate that \sysname{} can reliably converge to developer-aligned repairs even on vulnerabilities unseen during training.

\begin{tcolorbox} [boxrule=0.8pt,
                top=0.2pt,
                  bottom=0.2pt]
    \textbf{Answer to RQ2}: \sysname{} successfully repaired unseen vulnerabilities in the SAN2VULN benchmark, producing root-cause fixes in most cases and demonstrating generalization beyond its training data.
\end{tcolorbox}

\subsection{Repairing Vulnerabilities with Fixes Away from the Crash Site (RQ3)}
\label{ss:rq3}

\begin{table}[t]
\centering
\small
\caption{SEC-bench comparison on the 200 CVE evaluation samples. \emph{Resolved} reports acceptance counts (Criteria: Medium) for final patches generated by each backend. \emph{File} and \emph{Function} report localization coverage (\%) of the DeepSeek-V4.1-Flash patches against the developer patches. The second and third blocks restrict the evaluation to instances whose developer fix does not modify the crash-site function. $^\dagger$Reported in the original paper.}
\label{tab:secbench}
\newcolumntype{C}[1]{>{\centering\arraybackslash}p{#1}}

\begin{tabular}{@{}lccC{2.6cm}C{2.6cm}@{}}
\toprule
 & \multicolumn{2}{c}{Resolved (\%)} & \multicolumn{2}{c}{Localization (DeepSeek-V4.1-Flash)} \\
\cmidrule(lr){2-3}\cmidrule(l){4-5}
Method & GPT-5-mini & DeepSeek-V4.1-Flash & File (\%) & Function (\%) \\
\midrule
\multicolumn{5}{@{}l}{\emph{All instances (200)}} \\
SAN2PATCH~\cite{kim2025logs}       & 156 (78.0) & 147 (73.5) & 61.6 & 51.0 \\
PatchAgent~\cite{yu2025patchagent} & 168 (84.0) & 175 (87.5) & 65.5 & 56.8 \\
DebugHarness~\cite{debugharness}   & 141 (70.5) & 175 (87.5) & 65.5 & 58.7 \\
EvoRepair~\cite{evorepair}         & 174$^\dagger$ (87.0) & 186 (93.0) & 68.6 & 63.7 \\
\sysname{}                         & \textbf{178 (89.0)} & \textbf{195 (97.5)} & \textbf{71.8} & \textbf{66.0} \\
\midrule
\multicolumn{5}{@{}l}{\emph{All instances with a non-crash-site fix (112)}} \\
SAN2PATCH~\cite{kim2025logs}       & 83 (74.1) & 77 (68.8)  & 53.1 & 37.3 \\
PatchAgent~\cite{yu2025patchagent} & 90 (80.4) & 93 (83.0)  & 56.5 & 47.3 \\
DebugHarness~\cite{debugharness}   & 69 (61.6) & 95 (84.8)  & 57.8 & 50.7 \\
EvoRepair~\cite{evorepair}         & --        & 103 (92.0) & 61.2 & 57.3 \\
\sysname{}                         & \textbf{97 (86.6)} & \textbf{111 (99.1)} & \textbf{65.8} & \textbf{61.3} \\
\midrule
\multicolumn{5}{@{}l}{\emph{Excluding instances whose bug description names the fixed function (85)}} \\
SAN2PATCH~\cite{kim2025logs}       & 62 (72.9) & 59 (69.4) & 49.6 & 35.1 \\
PatchAgent~\cite{yu2025patchagent} & 69 (81.2) & 71 (83.5) & 53.9 & 43.9 \\
DebugHarness~\cite{debugharness}   & 53 (62.4) & 71 (83.5) & 52.2 & 43.9 \\
EvoRepair~\cite{evorepair}         & --        & 79 (92.9) & 56.5 & 51.8 \\
\sysname{}                         & \textbf{75 (88.2)} & \textbf{84 (98.8)} & \textbf{61.7} & \textbf{59.6} \\
\bottomrule
\end{tabular}
\end{table}

To examine this research question, we use the 200-CVE evaluation split of SEC-bench, a larger benchmark in which 112 instances (56\%) have a developer fix that does not modify the crash-site function. For these instances, the sanitizer report points to where the vulnerability manifests rather than where it should be repaired. In addition to SAN2PATCH and PatchAgent, which generated the most plausible patches among the baselines on VulnLoc (35 of 39 each), we include two recent agent-based AVR methods that were previously evaluated on SEC-bench, DebugHarness~\cite{debugharness} and EvoRepair~\cite{evorepair}. We run every method with both GPT-5-mini and DeepSeek-V4.1-Flash on the same instance set, except EvoRepair with GPT-5-mini, for which we take the result reported in the original paper; since only the aggregate count is reported there, its subset results are unavailable. For EvoRepair with DeepSeek-V4.1-Flash, we run up to six turns, after which no additional instances were newly resolved. \autoref{tab:secbench} reports the results on all instances and on the non-crash-site subset. Since EvoRepair also takes the CVE description provided by SEC-bench as input, which names a developer-modified function in 27 of the 112 instances and may thus reveal the fix location, we additionally report the results on the remaining 85 instances.

\paragraph{Repair results.}
On all 200 instances, \sysname{} resolves 178 and 195 instances with GPT-5-mini and DeepSeek-V4.1-Flash, respectively, the most among all methods under both backends. The advantage persists on the 112 non-crash-site instances. With DeepSeek-V4.1-Flash, \sysname{} resolves 111 of these 112 instances (99.1\%), whereas the resolution rates of all baselines decrease relative to the full set: SAN2PATCH from 73.5\% to 68.8\%, PatchAgent from 87.5\% to 83.0\%, DebugHarness from 87.5\% to 84.8\%, and EvoRepair from 93.0\% to 92.0\%. With GPT-5-mini, \sysname{} shows the smallest drop (89.0\% to 86.6\%), compared with 3.6 to 8.9 percentage points for SAN2PATCH, PatchAgent, and DebugHarness. On the 85 instances whose description does not name a fixed function, \sysname{} still resolves 84 with DeepSeek-V4.1-Flash and 75 with GPT-5-mini, compared with 79 and 69 for the strongest baselines under each backend.

\paragraph{Localization.}
Because SEC-bench acceptance is execution-based, resolution alone does not show whether a patch modifies the locations changed by the developer. We therefore also evaluate the localization of the final patches generated with DeepSeek-V4.1-Flash across all 200 instances. Function-level results cover 198 instances, as the developer patches for the remaining two modify only a macro or a grammar file. We adapt EvoRepair's localization evaluation protocol~\cite{evorepair}; metric definitions and protocol modifications are detailed in \autoref{sec:app_localization}.
On the full set, \sysname{} achieves the highest function-level coverage (66.0\%), 2.3 percentage points above EvoRepair (63.7\%). Moving to the non-crash-site subset, function-level coverage drops for every method, but least for \sysname{}: by 4.7 points, compared with 6.4 (EvoRepair), 8.0 (DebugHarness), 9.5 (PatchAgent), and 13.7 (SAN2PATCH). On the 85 remaining instances, the gap over EvoRepair widens to 7.8 points (59.6\% vs.\ 51.8\%), and \sysname{}'s coverage is 15.7 to 24.5 points higher than the other baselines. These results are consistent with our design: baselines that anchor localization on the crash-site frame lose accuracy when that frame is not the repair target, whereas patch-conditioned re-localization and root-cause-oriented patch review let \sysname{} move the fix toward the developer's location.

\begin{tcolorbox} [boxrule=0.8pt, top=0.2pt, bottom=0.2pt]
\textbf{Answer to RQ3:} On SEC-bench, \sysname{} resolves 111 of 112 instances whose fix lies outside the crash-site function, and its function-level localization coverage drops the least when the crash site is not the repair target, widening its lead over the strongest baseline from 2.3 to 7.8 percentage points once function-name hints are removed.
\end{tcolorbox}

\subsection{Impact of Specialized Agents on Repair Performance (RQ4)} \label{ss:rq4}

\begin{table}[t]
\centering
\caption{Impact of specialized-agent ablation and role merging on AVR performance.}
\label{tab:rq3}
\newcolumntype{C}[1]{>{\centering\arraybackslash}p{#1}}

\begin{subtable}{1\textwidth}
\centering
\scriptsize
\setlength{\tabcolsep}{10pt}
\renewcommand{\arraystretch}{1.2}
\begin{tabular}{ccc}
\hline
Setting* & Plausible Patches & Correct Patches \\
\hline
\makecell{\textbf{\sysname{}}} & 39 & 36 \\
\hline
\makecell{No CA} & 32 (-7) & 30 (-6) \\
\makecell{No SA} & 31 (-8) & 29 (-7) \\
\makecell{No FA} & 32 (-7) & 29 (-7) \\
\makecell{No RA} & 39 (-0) & 32 (-4) \\
\hline
\makecell{SA (w/o PCRL)} & 36 (-3) & 34 (-2) \\
\hline
\end{tabular}
\caption{Component ablation: effect of removing individual agents and disabling Patch-conditioned Re-localization.}
\label{tab:ablation_agents}
\end{subtable}%

\begin{subtable}{1\textwidth}
\centering
\scriptsize
\setlength{\tabcolsep}{10pt}
\renewcommand{\arraystretch}{1.2}
\begin{tabular}{ccc}
\hline
Setting* & Plausible Patches & Correct Patches \\
\hline
\makecell{\textbf{\sysname{}}} & 39 & 36 \\
\hline
\makecell{FLA + PGA} & 31 (-8) & 27 (-9) \\
\makecell{SA + FA} & 33 (-6) & 32 (-4) \\
\makecell{CA + FA} & 35 (-4) & 34 (-2) \\
\hline
\end{tabular}
\caption{Agent merging: effect of merging the roles of two agents into a single module.}
\label{tab:merge_agents}
 \end{subtable}%
 
\footnotesize
\begin{flushleft}
*FLA: Fault Localization Agent, PGA: Patch Generation Agent, CA: Compile Validation Agent, SA: Security Validation Agent, FA: Functionality Validation Agent, RA: Patch Review Agent, PCRL: Patch-conditioned Re-localization.
\end{flushleft}

\end{table}

To answer this research question, we conduct a component ablation study on \sysname{}’s specialized agents and its Patch-conditioned Re-localization (PCRL) capability. As shown in~\autoref{tab:ablation_agents}, we remove each component in turn and report the resulting numbers of plausible and correct patches. For the Compile Validation Agent (CA), Security Validation Agent (SA), and Functionality Validation Agent (FA), the removed agent is replaced with a simple tool that returns only validation outputs, without agent-guided feedback. We also evaluate a variant that retains the Security Validation Agent but disables PCRL.
The results show that all four specialized agents, as well as PCRL, contribute to repair performance. Removing CA, SA, and FA reduces the number of plausible patches by 7, 8, and 7, respectively, while disabling PCRL reduces it by 3. Removing CA, SA, FA, and the Patch Review Agent (RA) reduces the number of correct patches by 6, 7, 7, and 4, respectively, and disabling PCRL reduces it by 2. These results indicate that the agents improve repair not only through validation, but also by providing actionable feedback.
We also observed that excluding certain agents often caused the system to reach the maximum iteration limit and incur substantial computational costs.

Next, \autoref{tab:merge_agents} evaluates the impact of merging two agents into a single role by combining their prompts. Overall, role merging consistently degrades performance, indicating the importance of specialization even with identical tool access.
The largest drop occurs when merging Fault Localization and Patch Generation (FLA+PGA), reducing plausible patches from 39 to 31 and correct patches from 36 to 27 (8 and 9 additional failures). Merging validation roles has a smaller but still negative effect: SA+FA yields 33 plausible and 32 correct patches (6 and 4 additional failures), while CA+FA results in 35 plausible and 34 correct patches (4 and 2 additional failures).
These results suggest that separating localization, generation, and validation enables more targeted feedback and improves convergence to correct patches.

\begin{tcolorbox}[boxrule=0.8pt,
                  top=0.2pt,
                  bottom=0.2pt]
    \textbf{Answer to RQ4}: Each component in \sysname{} is indispensable, and its collaboration with the others is essential for generating optimal patches. 
    The Patch Review Agent further contributes by producing more complete patches.
\end{tcolorbox}

\begin{table}[t]
\tiny
\centering
\caption{Average token usage and API cost per vulnerability of AVR tools in the experiments described in \autoref{ss:rq1}. API costs are reported in USD.}
\label{tab:rq4}
\resizebox{0.65\linewidth}{!}{%
\begin{tabular}{c c c c c c}
\hline
Approach
& \makecell[c]{Input\\[-1pt]Tokens}
& \makecell[c]{Cached\\[-1pt]Tokens}
& \makecell[c]{Output\\[-1pt]Tokens}
& \makecell[c]{API Cost\\[-1pt](\$)}
& \makecell[c]{Runtime\\[-1pt](min)}
\\
\hline
\makecell{CrashRepair~\cite{crashrepair}} & \makecell{-} & \makecell{-} & \makecell{-} & \makecell{-} & \makecell{11.13}
\\
\makecell{LoopRepair~\cite{looprepair}} & \makecell{82,177} & \makecell{53,066} & \makecell{50,734} & \makecell{0.49} & \makecell{99.64}
\\
\makecell{SAN2PATCH~\cite{kim2025logs}} & \makecell{183,951} & \makecell{68,516} & \makecell{24,759} & \makecell{0.46} & \makecell{13.34}
\\
\makecell{PatchAgent~\cite{yu2025patchagent}} & \makecell{1,063,448} & \makecell{870,357} & \makecell{15,543} & \makecell{0.95} & \makecell{11.13}
\\
\makecell{SpecRover~\cite{specrover}} & \makecell{1,747,280} & \makecell{1,182,950} & \makecell{49,198} & \makecell{2.11} & \makecell{27.05}
\\
\makecell{\sysname{}} & \makecell{1,981,518} & \makecell{1,634,846} & \makecell{16,232} & \makecell{1.64} & \makecell{15.17}
\\
\hline
\end{tabular}
}
\end{table}

\subsection{Cost and Runtime (RQ5)} \label{ss:rq5}
In this section, we report per-sample API cost and end-to-end runtime for the experiments in \autoref{ss:rq1}. 
\autoref{tab:rq4} summarizes token usage, estimated API cost (USD), and end-to-end runtime for each approach. 
All metrics are averaged over 39 samples.
\sysname{} consumes the largest number of input and cached tokens (1.98M input, 1.63M cached), which reflects the overhead of multi-agent coordination, repeated validation, and patch-conditioned feedback across iterations. However, its output tokens remain small (16.2K), since patches are short and most tokens are spent on analysis and tool-guided reasoning rather than verbose generation.

In terms of cost, \sysname{} (\$1.64) is more expensive than LoopRepair (\$0.49), SAN2PATCH (\$0.46), and PatchAgent (\$0.95),  but cheaper than SpecRover (\$2.11). SpecRover and \sysname{} incur higher token usage primarily due to their multi-agent workflows. For runtime, \sysname{} completes in 15.17 minutes on average, which is close to SAN2PATCH (13.34 min) and slower than CrashRepair and PatchAgent (11.13 min), while being substantially faster than LoopRepair (99.64 min) and SpecRover (27.05 min). Overall, these results suggest that \sysname{} trades additional token cost for stronger validation and feedback, while maintaining practical runtime comparable to the fastest end-to-end baselines. The maintenance cost caused by an incorrect patch includes service downtime and long-term system instability, thereby increasing maintenance overhead. In comparison, an additional \$1 API cost for pre-deployment validation is a trivial investment that is far more cost-effective.

\begin{tcolorbox}[boxrule=0.8pt,
                  top=0.2pt,
                  bottom=0.2pt]
\textbf{Answer to RQ5}: \sysname{} incurs higher token cost than some baselines, but delivers more developer-aligned repairs, with fewer PoC-only fixes, while maintaining end-to-end runtime comparable to the fastest systems.
\end{tcolorbox}

\section{Discussion}

\subsection{Sensitivity to the LLM Backend}
\label{app:llm_backend_sensitivity}

\begin{table}[!t]
    \centering
    \caption{Total effectiveness and cost for different LLMs.}
    \label{tab:llm_backend_swap_vulnloc}
    \renewcommand{\arraystretch}{0.92}
    \scriptsize
\begin{tabular}{lccccc}
\hline
\makecell[c]{LLM Configuration} & 
\makecell[c]{Plausible \\ Patches} & 
\makecell[c]{Input \\ Tokens} &
\makecell[c]{Output \\ Tokens} &
\makecell{API Cost \\ (\$)}
\\
\hline
Gemini-2.5-pro~\cite{geminipro} & 39 (-0) & 4,203,009 & 95,888 & 2.29 \\
Claude-haiku-4.5~\cite{claudehaiku} & 34 (-5) & 3,488,775 & 36,962 & 3.67 \\
\hline
\end{tabular}
\end{table}

To assess \sysname{}'s sensitivity to the LLM backend, we swap GPT-4.1 with Gemini-2.5-Pro~\cite{geminipro} and Claude Haiku 4.5~\cite{claudehaiku} while keeping the multi-agent workflow, tool access, loop budget, and validation pipeline fixed. Table~\ref{tab:llm_backend_swap_vulnloc} summarizes the results. \sysname{} remains effective across backends: Gemini-2.5-Pro generates plausible patches for all 39 VulnLoc samples, while Claude Haiku 4.5 generates 34 out of 39. This suggests that much of \sysname{}'s effectiveness comes from the workflow design, although backend capability still affects the final repair rate.

We also observe clear differences in token usage and cost. Relative to the default setting, Gemini-2.5-Pro increases the average cost from \$1.44 to \$2.29, largely because it tends to invoke more tools before producing a final repair decision. Claude Haiku 4.5 raises the average cost further to \$3.67, even though it is intended as a fast and relatively low-cost model, because caching was difficult to use effectively in our setup. These results show that backend-specific interaction patterns can significantly affect token efficiency, and that further prompt and tool optimization could reduce cost.

\subsection{Case Studies}
\label{s:CaseStudy}

\providecommand{\textpanel}[1]{%
  \par\noindent
  \fcolorbox{dvFrame}{dvBg}{%
    \parbox{\dimexpr\linewidth-2\fboxsep-2\fboxrule\relax}{%
      \scriptsize\raggedright #1}}%
  \par}
\providecommand{\dvitem}[2]{%
  \par\hangindent=1.1em\hangafter=1\noindent
  \hspace*{0.3em}\makebox[0.8em][l]{\textbullet}\textbf{#1:} #2}

\begin{figure*}[ht!]
    \centering
    \begin{subfigure}[t]{\linewidth}
        \codehdr{Patch Location Candidates (ranked) \hfill First analysis}
        \textpanel{%
            \textbf{1. \textcolor{dvRem}{src/util/parser.c (line range: 973--1011)} [Confidence: HIGH]}
            \dvitem{Evidence}{This is where \texttt{act->Params} is repeatedly \texttt{realloc}'d, invalidating pointers held elsewhere.}
            \dvitem{Fix Strategy}{Change the allocation strategy to avoid reallocating and moving the array after each parameter. Instead, allocate the full array up front if possible, or use a temporary list and copy at the end. Alternatively, store all parameters in a list and only assign to \texttt{act->Params} after all are parsed.}}

        \vspace{0.3em}

        \codehdr{Patch Location Candidates (ranked) \hfill Final analysis}
        \textpanel{%
            \textbf{1. \textcolor{dvAdd}{src/util/decompile.c (line range: 348, 361)} [Confidence: HIGH]}
            \dvitem{Evidence}{If \texttt{act->p.Constant8} or \texttt{act->p.Constant16} is out of bounds, any access to \texttt{pool[...]} will crash, regardless of NULL checks.}
            \dvitem{Fix Strategy}{Add a check that \texttt{pool} is non-NULL and that \texttt{act->p.Constant8} (or \texttt{Constant16}) is within the valid range (pool size) before any use in \texttt{getString} and \texttt{getName}. If not, return a static safe string (\texttt{"(null)"}).}}
        \captionsetup{margin=0.3em}
        \caption{(CVE-2018-8806) Initial vs.\ final top-ranked fault-localization candidates of \sysname.}
        \label{subfig:fl_8806}
    \end{subfigure}

    \vspace{0.6em}

    \begin{subfigure}[t]{0.63\linewidth}
        \codehdr{/src/libjasper/jpc/jpc\_cs.c \hfill jpc\_siz\_getparms}
\begin{lstlisting}[style=dvDiff]
 /* developer fix */
    jas_free(siz->comps); return -1; }
+ if (siz->comps[i].hsamp == 0 || siz->comps[i].hsamp > 255) {
+   jas_eprintf("invalid XRsiz value %d\n", siz->comps[i].hsamp);
+   jas_free(siz->comps); return -1; }
+ if (siz->comps[i].vsamp == 0 || siz->comps[i].vsamp > 255) {
+   jas_eprintf("invalid YRsiz value %d\n", siz->comps[i].vsamp);
+   jas_free(siz->comps); return -1; }
  siz->comps[i].sgnd = (tmp >> 7) & 1;
  siz->comps[i].prec = (tmp & 0x7f) + 1;

 /* DeepAVR */
    jas_free(siz->comps); return -1; }
+ if (siz->comps[i].hsamp == 0 || siz->comps[i].vsamp == 0) {
+   jas_free(siz->comps); return -1; }
  siz->comps[i].sgnd = (tmp >> 7) & 1;
  siz->comps[i].prec = (tmp & 0x7f) + 1;
\end{lstlisting}
        \captionsetup{margin=0.3em}
        \caption{(CVE-2016-8691) Correct \sysname{} patch.}
        \label{subfig:patch_8691}
    \end{subfigure}
    \hfill
    \begin{subfigure}[t]{0.35\linewidth}
        \codehdr{/src/shred.c \hfill fillpattern}
\begin{lstlisting}[style=dvDiff]
 /* developer fix */
  r[0] = (bits >> 4) & 255;
  r[1] = (bits >> 8) & 255;
  r[2] = bits & 255;
- for (i = 3; i < size / 2; i *= 2)
+ for (i = 3; i <= size / 2; i *= 2)
    memcpy (r + i, r, i);
  if (i < size)
    memcpy (r + i, r, size - i);

 /* DeepAVR */
  for (i = 3; i < size / 2; i *= 2)
-   memcpy (r + i, r, i);
+   memmove (r + i, r, i);
  if (i < size)
-   memcpy (r + i, r, size - i);
+   memmove (r + i, r, size - i);
\end{lstlisting}
        \captionsetup{margin=0.3em}
        \caption{(gnubug-26545) Plausible but incorrect \sysname{} patch.}
        \label{subfig:patch_26545}
    \end{subfigure}

    \caption{\sysname{} case studies on logically distant, physically distant, and failure scenarios.}
    \label{Fig:case_study}
\end{figure*}

This section presents three case studies: two that \sysname{} successfully repairs, and one that \sysname{} does not.

\PP{Resolving a Logically Distant Vulnerability (CVE-2018-8806).} 
The patch generation process for this example highlights both its effectiveness and the need for \sysname{} due to logical distance.
CVE-2018-8806 is a use-after-free bug in \texttt{libming}~\cite{libming}, triggered when the array \texttt{pool} is accessed using \texttt{act->p.Constant8} as the index. The sanitizer log for this vulnerability reports a use-after-free (UAF) similar to \autoref{subfig:san_8964}. However, the actual sequence is that an out-of-bounds access on \texttt{pool} overwrites a neighboring freed region, which the sanitizer flags as a UAF. As shown in \autoref{subfig:fl_8806}, the Fault Localization Agent first selects a candidate in \texttt{parser.c} associated with pointer invalidation, assuming the root cause is UAF. Once a patch is generated for this site, the sanitizer instead reports a null pointer dereference. Due to patch-conditioned fault localization, the Fault Localization Agent eventually converges on a final candidate in \texttt{decompile.c} (lines 348–361), identified as the true root cause. The analysis indicates that if \texttt{act->p.Constant8} or \texttt{act->p.Constant16} is out of bounds, any access to \texttt{pool[...]} will result in a crash regardless of null checks.

\PP{Resolving a Physically Distant Vulnerability (CVE-2016-8691).}
CVE-2016-8691 is a divide-by-zero bug in Jasper caused by missing validation of \texttt{cmpt->hstep} and \texttt{cmpt->vstep}. Because the optimal location for the check is physically distant from the crash site, localized fixes at \texttt{jpc\_dec.c} can suppress the immediate symptom without fixing the root cause. DeepAVR initially produced such a crash-site guard, but the Patch Review Agent identified it as a \textit{symptom-level fix} and recommended moving the validation to the SIZ parsing logic. Guided by this feedback, DeepAVR relocated the fix to \texttt{jpc\_cs.c}, the critical file for the repair but one absent from the sanitizer log, and generated a patch similar to the developer fix (\autoref{subfig:patch_8691}). Although our patch omits the developer's upper-bound checks for values \(>255\), the relevant parameter type (\texttt{uint\_fast8\_t}, implemented here as \texttt{unsigned char}) makes overflow negligible, so the patch still correctly fixes the reported divide-by-zero root cause.

\PP{Failure Case of \sysname{} (gnubug-26545).}
This example is a \texttt{memcpy}-param-overlap bug in \texttt{coreutils}~\cite{coreutils} triggered when the source and destination regions overlap, causing undefined behavior. Although all patches \sysname{} generated pass the project’s security and functionality tests, manual analysis revealed that they alter functionality.
The developer’s patch (\autoref{subfig:patch_26545}) adjusts the loop condition to avoid overlap, whereas every LLM-generated patch, including ours, replaces nearby \texttt{memcpy} calls with \texttt{memmove}.
Though this prevents undefined behavior, it conflicts with the developer’s intent.
This case exposes the limits of the project’s testing framework, underscores the need for a more comprehensive test suite for the project, and reveals a \sysname{} limitation bounded by the test suite's completeness.

\subsection{Threats to Validity}

\PP{Internal Validity.}  
For the VulnLoc and SAN2VULN evaluations, we replicated the LLM-based baselines using their released artifacts and replaced their original base models with GPT-4.1 to standardize the backend. For SEC-bench, we evaluated methods with GPT-5-mini and DeepSeek-V4.1-Flash, except that EvoRepair's
GPT-5-mini aggregate result was taken from its original paper.
Changing the baselines' original models may have affected their efficiency and repair performance.
Some methods also rely on external localization components. In particular, LoopRepair~\cite{looprepair} depends on CrashAnalysis~\cite{crashrepair} for iterative localization. In our environment, some samples excluded in the original LoopRepair evaluation due to failed localization were initially localizable, but re-localization sometimes failed after candidate patches were applied, which may have negatively affected their performance in our setting. 

For VulnLoc and SAN2VULN, patch correctness was assessed primarily through automated validation (security and functionality tests), complemented by manual inspection. To mitigate author bias, manual evaluation was conducted by three Ph.D.\ 
students who are not authors of this paper, each independently reviewing 
patches without knowledge of which tool produced them. Reviewers first identified the violated invariant from the sanitizer report, the execution path derived from the log after executing the PoC, the vulnerable code, the bug type, and any available external evidence, such as the developer's patch, if accessible. We define an invariant as the condition that must hold before the vulnerability is triggered to prevent the reported violation while
preserving intended behavior. A patch was labeled correct only if it
enforced this invariant at the semantically responsible point, covered
equivalent risky code paths sharing the same invariant, and preserved
intended functionality. Reviewers independently labeled
the anonymized patches using this rubric, and disagreements were
resolved through rubric-based discussion without revealing tool
identity.

\PP{External Validity.}  
All evaluations were performed on C/C++ vulnerabilities from the VulnLoc, SAN2VULN, and SEC-bench benchmarks.
Thus, the results may not generalize to other programming languages or codebases. 
Furthermore, \sysname{}’s performance depends on the choice of LLM backend (\autoref{app:llm_backend_sensitivity}), and outcomes may differ under alternative models or future model versions.

\section{Related Work} \label{s:BG}

\PP{Automated Vulnerability Repair (AVR).}
Several APR studies have specialized in repairing security vulnerabilities~\cite{huang2024evolving}. Analysis-based AVR systems relax the need for full test suites by repairing from a single failing input, but often rely on stronger assumptions: SenX~\cite{senx} requires human-specified safety properties, CPR~\cite{cpr} and VulnFix~\cite{zhang2022program} assume known fix locations, and ExtractFix~\cite{extractfix} relies on explicit crash constraints. In contrast, CrashRepair~\cite{crashrepair} removes these assumptions and instead repairs from only a single PoC using concolic execution, constraint inference, mutation-based patch generation, and differential testing. 
EffFix~\cite{zhang2025efffix} repairs memory errors using feedback from incorrectness-separation-logic-based static analysis to guide patch search, and reduces validation cost by grouping patches with equivalent
effects on the symbolic heap.
Our work instead investigates how specialized LLM agents interpret intermediate repair outcomes to revise localization and identify incomplete repairs.

\PP{LLM-based AVR.}
Early LLM-based AVR studies assume perfect FL, where the buggy function is directly provided~\cite{sp23,sec24,AVR_ASE25_Vul-R2,AVR_ASE25_MUSSEL,AVR_TOSEM24_VQM}. 
In this setting, fine-tuning-based approaches~\cite{icse24} learn repair patterns from paired buggy--fixed functions; for example, VulRepair~\cite{fse22} fine-tunes T5 for vulnerability repair. In parallel, prompt-engineering approaches have demonstrated that general-purpose LLMs can also repair vulnerabilities without task-specific training~\cite{sp23,sec24}.

More recently, end-to-end AVR systems have begun integrating localization, patch generation, and validation directly from weak bug reports such as PoCs or sanitizer logs~\cite{zhang2026fixing}. Representative examples include LoopRepair~\cite{looprepair}, which combines CrashAnalysis~\cite{crashrepair} with iterative repair, SAN2PATCH~\cite{kim2025logs}, which decomposes repair into staged LLM reasoning, PatchAgent~\cite{yu2025patchagent}, which frames repair as tool-augmented debugging, and ATLANTIS-Patching~\cite{kim2025atlantis}, which improves robustness through agent ensembling. DebugHarness~\cite{debugharness} supports interactive inspection of runtime states and execution flow to guide vulnerability repair. EvoRepair~\cite{evorepair} accumulates and retrieves repair experience within and across vulnerability instances. \sysname{} focuses on failure-aware feedback routing among specialized agents, patch-conditioned re-localization, and review of test-passing patches for symptom-level fixes and incomplete repairs.

\PP{LLM for Program Repair.} %
The use of LLMs in program repair has gained significant attention in recent years~\cite{ yang2024sweagent, ouyang2025repograph, antoniades2025swesearch, zhang2024autocoderover, APR_icse26_Reinfix, APR_icse25_D4C, APR_icse25_NTR}. 
A recent and influential direction is \emph{iterative repair}, where candidate patches are repeatedly refined using feedback from prior repair attempts. In feedback-driven iterative approaches, the repair process is often guided either by lightweight validation signals, such as the number or ratio of passing tests~\cite{APR_icse26_Flames, ye2024iter}, or by agent-analyzed feedback, where an LLM inspects failed tests, error messages, or other execution artifacts to refine subsequent patch candidates~\cite{APR_icse26_MGDebugger, APR_ISSTA24_ChatRepair, APR_ISSTA24_thinkrepair, bouzenia2025repairagent}.
Despite their effectiveness, most approaches focus on iterative patch generation while treating FL as fixed, either assuming perfect localization or relying on existing tools without actively revisiting it during repair.
Although ITER~\cite{ye2024iter} integrates localization, generation, and validation in an end-to-end formulation, their refinement still depends on test-driven signals, which are often weak or unavailable in vulnerability repair scenarios where only a single failing test case (e.g., a PoC) is provided.

Another line of work explores \emph{multi-agent collaboration} for software engineering tasks, such as test-driven debugging~\cite{lee2025unidebugger} and GitHub issue resolution~\cite{tao2024magis, specrover}. These systems show that decomposing complex workflows into specialized roles can improve reasoning, coordination, and review beyond single-agent settings. In repair-oriented systems, this often takes the form of reviewer-style feedback. For example, SpecRover~\cite{specrover} employs a dedicated reviewer agent that analyzes generated patches and provides natural-language feedback for subsequent iterations. This goes beyond simple test-based refinement or single-agent self-reflection, but the feedback remains only partially coupled; it refines patch generation without being propagated back to the localization stage. More generally, existing approaches do not explicitly model the diversity of repair failure modes. They rarely distinguish symptom-level fixes from root-cause corrections, reason about newly introduced vulnerabilities, or decompose feedback into dimensions such as localization errors, broken functionality, and security flaws.

\section{Conclusion}
This paper presented \sysname{}, a multi-agent framework for end-to-end automated vulnerability repair. By decomposing repair into specialized roles, refining hypotheses through structured feedback, and leveraging patch-conditioned re-localization, \sysname{} supports deeper analysis under weak initial signals. It further uses patch review to identify superficially valid but semantically incorrect fixes.
Experiments on real-world benchmarks show that \sysname{} outperforms prior end-to-end AVR systems and reduces the number of plausible but incorrect fixes. Overall, our results suggest that structured agent collaboration is a promising direction for robust automated vulnerability repair.

\section*{Ethics and Privacy Statement}

This study uses publicly available vulnerability-repair benchmarks and open-source software projects and does not involve human subjects or personal data. \sysname{} is intended to assist in repairing software vulnerabilities by generating and validating candidate patches. Its outputs are not guaranteed to be correct or secure and should therefore undergo conventional testing, static and dynamic analysis, and expert review before deployment. Because the framework operates on vulnerable software and proof-of-concept inputs, it should be used only on authorized codebases and in controlled environments.

\bibliographystyle{ACM-Reference-Format}
\bibliography{References}

\appendix

\section{Agent Prompts}
\label{app:prompts}
This section provides condensed versions of the system prompts used by the agents of \sysname{}.
All agents share a common prompt skeleton that specifies the agent's role, available tools and their usage, key instructions, expected inputs, and output schema; the figures retain only the instructions that distinguish each agent, and each box uses the color of the corresponding agent in \autoref{fig:overview}.

\autoref{fig:prompt_FL} shows the prompt of the Fault Localization Agent. Rather than stopping at the crash site, it walks the call stack in reverse and explores sibling calls that execute before the crash, since their side effects often create the invalid state. It maintains a ranked candidate queue across iterations and re-proposes a candidate only when new feedback changes the evidence.
\autoref{fig:prompt_PG} shows the prompt of the Patch Generation Agent, which prioritizes root-cause edits, preserves existing semantics, and re-reads the target lines immediately before emitting a patch to avoid hallucinated context.
\autoref{fig:prompt_CV} shows the prompt of the Compile Validation Agent, which is restricted to applying and building patches and classifies each failure so that the Patch Generation Agent can revise precisely.
\autoref{fig:prompt_SV} shows the prompt of the Security Validation Agent. It compares only normalized bug types, so that a changed type triggers patch-conditioned re-localization (\autoref{ss:patch-cond}); for a non-zero exit without a sanitizer report, it accepts only an orderly refusal of the malicious input.
\autoref{fig:prompt_FV} shows the prompt of the Functionality Validation Agent, which combines a strict test verdict with a manual impact analysis on valid inputs.
The prompt of the Patch Review Agent is shown in \autoref{fig:prompt_PR} and discussed in \autoref{ss:Patch_Review}.

\definecolor{flBorder}{HTML}{D7A9DB}
\begin{figure}[h]
\centering
\setlength{\fboxsep}{0pt}
\setlength{\fboxrule}{1pt}
\fcolorbox{flBorder}{white}{%
\begin{minipage}{0.9\linewidth}
{\setlength{\fboxsep}{2pt}\colorbox{flBorder}{\parbox{\dimexpr\linewidth-4pt}{\centering\small\sffamily\bfseries Fault Localization Agent}}}\par\vspace{3pt}
\hspace*{5pt}\begin{minipage}{\dimexpr\linewidth-10pt}
\footnotesize\sffamily\raggedright
Locate the \emph{root cause} of the sanitizer-reported crash, not merely the crash site:\\[1pt]
\textbf{1) Object-Centric Reverse Call-Stack Traversal}\\
Identify the crash-related objects (pointers, arrays, struct fields) and walk the stack from the crash frame $F_0$ up to the outermost frame $F_k$.
In each frame, analyze the code from function entry to the call site of $F_{i-1}$, and explore \emph{all sibling calls} made before it (DFS, depth $\leq 3$),
since their side effects (e.g., header sizes, offsets) may set up the later crash.\\[1pt]
\textbf{2) Root Cause vs.\ Symptom}\\
Trace each object from allocation/initialization to failure and look for broken variable dependencies, violated invariants, and allocation--usage mismatches.
A read/write on already-corrupted data is a \emph{symptom}.
Rank fixes as: wrong allocation size $>$ missing synchronized update of related fields (e.g., length vs.\ capacity) $>$ inconsistent data-structure relationships $>$ validation checks (fallback only).\\[1pt]
\textbf{3) Candidate Queue Across Iterations}\\
Maintain a ranked queue of \{file, line range, confidence, fix type (\textsc{root cause}/\textsc{symptom}), rationale, status\}.
Promote exactly one untried candidate per iteration; re-propose a tried one only with an explicit \emph{evidence delta} from new feedback.\\[1pt]
\textbf{4) Exhaustion $\rightarrow$ Re-analysis $\rightarrow$ Expansion}\\
When no untried candidate remains, re-analyze the latest sanitizer report, revisiting skipped siblings and upstream producers of size/offset/count fields (parsers, header readers).
If still none, expand once by one caller level upstream and add at least two new root-cause candidates.
\end{minipage}\par\vspace{4pt}
\end{minipage}}
\caption{Tailored prompt for the Fault Localization Agent.}
\label{fig:prompt_FL}
\end{figure}

\definecolor{pgBorder}{HTML}{B1D095}
\begin{figure}[h]
\centering
\setlength{\fboxsep}{0pt}
\setlength{\fboxrule}{1pt}
\fcolorbox{pgBorder}{white}{%
\begin{minipage}{0.9\linewidth}
{\setlength{\fboxsep}{2pt}\colorbox{pgBorder}{\parbox{\dimexpr\linewidth-4pt}{\centering\small\sffamily\bfseries Patch Generation Agent}}}\par\vspace{3pt}
\hspace*{5pt}\begin{minipage}{\dimexpr\linewidth-10pt}
\footnotesize\sffamily\raggedright
Turn the selected candidate location and root-cause hypothesis into a minimal patch:\\[1pt]
\textbf{1) Root-Cause Priority}\\
Fix the origin of the vulnerability (allocation/size mismatch, violated invariant, faulty logic).
Fall back to a symptom-level guard only if the root cause cannot be safely patched in scope, and justify it explicitly.\\[1pt]
\textbf{2) Minimal, Semantics-Preserving Edits}\\
Change only what neutralizes the vulnerability.
Do not reorder calls, loops, or state updates, and keep existing error-handling conventions, unless they are themselves the fault source.\\[1pt]
\textbf{3) Grounded Emission}\\
Re-read the exact target lines with \texttt{FileReadLines} immediately before emitting each patch, and emit it as an
\emph{original code (line range)} / \emph{replacing code} pair.
Use identifiers already in scope and keep braces, control flow, and \texttt{\#if}/\texttt{\#endif} balanced.\\[1pt]
\textbf{4) Feedback-Driven Revision}\\
If security or review feedback reports an incomplete fix, move the edit toward a more fundamental point.
If regression tests fail, read the failing tests to recover the expected behavior.
Each new attempt must be a meaningfully improved revision, not a resubmission.
\end{minipage}\par\vspace{4pt}
\end{minipage}}
\caption{Tailored prompt for the Patch Generation Agent.}
\label{fig:prompt_PG}
\end{figure}

\definecolor{cvBorder}{HTML}{F9DA78}
\begin{figure}[h]
\centering
\setlength{\fboxsep}{0pt}
\setlength{\fboxrule}{1pt}
\fcolorbox{cvBorder}{white}{%
\begin{minipage}{0.9\linewidth}
{\setlength{\fboxsep}{2pt}\colorbox{cvBorder}{\parbox{\dimexpr\linewidth-4pt}{\centering\small\sffamily\bfseries Compile Validation Agent}}}\par\vspace{3pt}
\hspace*{5pt}\begin{minipage}{\dimexpr\linewidth-10pt}
\footnotesize\sffamily\raggedright
Apply the candidate patch and rebuild the project:\\[1pt]
\textbf{1) Least-Privilege Role}\\
Call \texttt{PatchCompile} first. Only apply and compile the given patch:
never generate, fix, or split a patch, and reject any patch that creates a new file.\\[1pt]
\textbf{2) Failure Classification}\\
On failure, report which case applies so the Patch Generation Agent can revise precisely:
(a) \emph{invalid patch format}, with the failing patch index;
(b) \emph{code mismatch}, with the actual content of the lines to be replaced;
(c) \emph{compile error}, with the full diagnostics, file paths, and line numbers.\\[1pt]
\textbf{3) Routing}\\
Start the reply with \texttt{SUCCESS\_COMPILE} (forward to security validation) or
\texttt{FAIL\_COMPILE} (return to patch generation).
\end{minipage}\par\vspace{4pt}
\end{minipage}}
\caption{Tailored prompt for the Compile Validation Agent.}
\label{fig:prompt_CV}
\end{figure}

\definecolor{svBorder}{HTML}{E9B48A}
\begin{figure}[h]
\centering
\setlength{\fboxsep}{0pt}
\setlength{\fboxrule}{1pt}
\fcolorbox{svBorder}{white}{%
\begin{minipage}{0.9\linewidth}
{\setlength{\fboxsep}{2pt}\colorbox{svBorder}{\parbox{\dimexpr\linewidth-4pt}{\centering\small\sffamily\bfseries Security Validation Agent}}}\par\vspace{3pt}
\hspace*{5pt}\begin{minipage}{\dimexpr\linewidth-10pt}
\footnotesize\sffamily\raggedright
Rerun the PoC once on the patched program and judge security only:\\[1pt]
\textbf{1) Bug-Type Normalization}\\
Normalize the sanitizer header to a bug type (e.g., \emph{heap-use-after-free}, \emph{null-dereference-read}).
Ignore shifts in file, line, or function; compare only the bug type with the original report.\\[1pt]
\textbf{2) Three-Way Verdict}\\
\emph{No crash} $\rightarrow$ \texttt{PASS\_EVAL\_SECURITY}.
\emph{Same bug type} $\rightarrow$ \texttt{FAIL\_EVAL\_SECURITY}.
\emph{Different bug type} $\rightarrow$ \texttt{BUG\_TYPE\_CHANGED}: keep the patch applied and trigger patch-conditioned re-localization.\\[1pt]
\textbf{3) Failure Diagnosis (Same Vul.)}\\
Give 1--3 evidence-backed hypotheses on why the patch did not block the crash, naming what localization or generation missed:
crash site vs.\ upstream root cause, missing edge-case guard, incomplete data-flow/aliasing reasoning, off-by-one or unit mismatch, or a related site needing a co-fix.\\[1pt]
\textbf{4) Non-Zero Exit Classification}\\
A non-zero exit without a sanitizer report is never accepted by default.
Classify it as \emph{orderly refusal} of the malicious input, \emph{over-restriction} (the patch rejects legitimate work),
\emph{infrastructure} failure, or \emph{unexplained}; only an orderly refusal may pass.
\end{minipage}\par\vspace{4pt}
\end{minipage}}
\caption{Tailored prompt for the Security Validation Agent.}
\label{fig:prompt_SV}
\end{figure}

\definecolor{fvBorder}{HTML}{93A9D8}
\begin{figure}[h]
\centering
\setlength{\fboxsep}{0pt}
\setlength{\fboxrule}{1pt}
\fcolorbox{fvBorder}{white}{%
\begin{minipage}{0.9\linewidth}
{\setlength{\fboxsep}{2pt}\colorbox{fvBorder}{\parbox{\dimexpr\linewidth-4pt}{\centering\small\sffamily\bfseries Functionality Validation Agent}}}\par\vspace{3pt}
\hspace*{5pt}\begin{minipage}{\dimexpr\linewidth-10pt}
\footnotesize\sffamily\raggedright
Check that the patch preserves the program's intended behavior:\\[1pt]
\textbf{1) Strict Test Verdict}\\
Pass only if at least one test ran and failures, errors, and crashes are all zero.
Zero executed tests, build/environment problems, or summaries that conflict with failure cues in the log count as failure;
when in doubt, choose the stricter reading.\\[1pt]
\textbf{2) Manual Impact Analysis}\\
Always simulate \emph{valid} inputs, distinct from the PoC, against the original and patched code, checking returns, state, I/O, boundaries, and invariants.
A behavior change fails the patch even if all tests pass; without a test suite, this analysis decides alone.\\[1pt]
\textbf{3) Scope}\\
Judge only valid-input behavior: security is out of scope, so a PoC crash reported as \texttt{BUG\_TYPE\_CHANGED} does not affect the verdict.\\[1pt]
\textbf{4) Actionable Feedback}\\
On \texttt{FAIL\_EVAL\_FUNCTIONALITY}, report the counters and representative failing tests,
and direct the Patch Generation Agent to inspect them before revising the patch.
\end{minipage}\par\vspace{4pt}
\end{minipage}}
\caption{Tailored prompt for the Functionality Validation Agent.}
\label{fig:prompt_FV}
\end{figure}

\subsection{Example Output of the Fault Localization Agent}
\label{app:fl_output}
\autoref{fig:fl_rca_output} shows how the Fault Localization Agent analyzes the root cause of CVE-2018-8964 (\autoref{Fig:motivation}) from the code base and the sanitizer report.
Although the sanitizer reports a use-after-free in \texttt{getName} in the decompiler, the agent traces the freed object back to the \texttt{realloc} of the parameter array in \texttt{parseSWF\_ACTIONRECORD} in the parser.
It ranks this parser site as a high-confidence root cause and the crash site as a low-confidence symptom, which is the localization that guides the repair.

\definecolor{flBorder}{HTML}{D7A9DB}
\definecolor{flHighlight}{HTML}{FCE38A}
\begin{figure}[t]
\centering
\setlength{\fboxsep}{5pt}
\setlength{\fboxrule}{1pt}
\fcolorbox{flBorder}{white}{%
\begin{minipage}{0.9\linewidth}
\footnotesize\sffamily\raggedright
\textbf{\#\# Error-occurring Objects}\\
struct SWF\_ACTIONPUSHPARAM * passed to getName (from pop() in decompile).\\
Heap region allocated in parseSWF\_ACTIONRECORD and later freed by realloc (old Params buffer).\\
Parameter fields used after free: act-{\textgreater}p.String, act-{\textgreater}p.Constant8, act-{\textgreater}p.Constant16.\\[4pt]
\textbf{\#\# Write Tracking Analysis}\\
ASAN path: use-after-free in getName (decompile.c:398), where it reads a SWF\_ACTIONPUSHPARAM *.\\
The freed memory is the old SWF\_ACTIONPUSHPARAM array resized via realloc in parseSWF\_ACTIONRECORD (parser.c:1008).\\
Key hazard: pointers to elements of the Params array escape (e.g., pushed onto a stack) and are later dereferenced after realloc potentially moves the array, making those pointers dangling.\\[4pt]
\textbf{\#\# Exploration Summary (Bounded DFS)}\\
Use site: getName $\leftarrow$ decompileDELETE $\leftarrow$ decompileAction(s) $\leftarrow$ outputSWF\_DOACTION $\leftarrow$ readMovie $\leftarrow$ main.\\
Free site: parseSWF\_ACTIONRECORD repeatedly grows the Params array with realloc while parsing.\\[4pt]
\textbf{\#\# Root Cause Analysis}\\
Dangling pointer via realloc: parseSWF\_ACTIONRECORD grows the SWF\_ACTIONPUSHPARAM array with realloc, which can relocate the buffer and free the old region.\\
Invariant violation: external references must not point into a buffer that may be reallocated. Here, decompiler-side stack entries hold pointers into the reallocated array.\\
Effect: pop() returns a pointer to an element that may have been invalidated by a later realloc, leading to use-after-free in getName.\\[4pt]
\textbf{\#\# Patch Location Candidates (ranked)}\\
{\setlength{\fboxsep}{1pt}\colorbox{flHighlight}{\parbox{\dimexpr\linewidth-2pt}{\raggedright ROOT CAUSE (HIGH): parser.c parseSWF\_ACTIONRECORD (around Params allocation/resizing, \#L964-1012).}}}\\
Fix by preventing pointer invalidation: allocate sufficient space up front, avoid storing external pointers into a resizable array, or redesign to store indices/copies instead of raw pointers.\\
SYMPTOM (LOW): decompile.c getName region (\#L373-420).\\
Defensive checks could reduce crashes but do not fix the upstream dangling-pointer creation.
\end{minipage}}
\caption{Concise version of the root cause analysis of Fault Localization Agent.}
\label{fig:fl_rca_output}
\end{figure}

\section{SpecRover Implementation}
\label{sec:app_specrover}
To adapt SpecRover for AVR, we made three changes. First, SpecRover’s GitHub-issue workflow typically uses an LLM-based reproducer agent to synthesize an executable test from informal “steps to reproduce,” since such steps are often incomplete or non-executable. In our setting, each fuzzing-found vulnerability already comes with a Proof-of-Vulnerability (PoV) input that reliably reproduces the crash in the fuzzing environment. We therefore disable the reproducer agent and directly use the provided PoV to trigger the failure. Second, because SpecRover’s retrieval and parsing logic is optimized for GitHub repositories and often tailored to Python, we modify its file filtering and function parsing to support C projects (C file discovery and C function extraction). Finally, we enforce regression testing as the final acceptance criterion and only report patches that pass the regression test suite.

\section{Patch Localization Evaluation}
\label{sec:app_localization}
To measure patch localization on SEC-bench, we adapt EvoRepair's evaluation protocol~\cite{evorepair}. File-level (function-level) localization measures the proportion of developer-modified files (functions) also modified by the generated patches, aggregated across all instances. We consider at most three developer-modified functions per instance.
We make two changes to this protocol. First, we use the official SEC-bench developer patches as ground truth, because 32 of the 200 developer patches distributed with EvoRepair contain build-environment changes (e.g., \texttt{configure}, \texttt{autogen.sh}) instead of the vulnerability fix. Second, instead of reading function names from diff hunk headers, which name the enclosing namespace, a \texttt{goto} label, or a preceding function for 25 developer patches, we map each changed line to its enclosing function by parsing the pre-patch source in each instance's Docker image.

\end{document}